\documentclass{aa}
\usepackage[varg]{txfonts}

\usepackage{natbib,twoopt}
\usepackage[colorlinks,linkcolor=black,urlcolor=blue,citecolor=black,breaklinks=true]{hyperref} 
\bibpunct{(}{)}{;}{a}{}{,}             
\makeatletter
  \newcommandtwoopt{\citeads}[3][][]{\href{http://adsabs.harvard.edu/abs/#3}%
    {\def\hyper@linkstart##1##2{}%
     \let\hyper@linkend\@empty\citealp[#1][#2]{#3}}}
  \newcommandtwoopt{\citepads}[3][][]{\href{http://adsabs.harvard.edu/abs/#3}%
    {\def\hyper@linkstart##1##2{}%
     \let\hyper@linkend\@empty\citep[#1][#2]{#3}}}
  \newcommandtwoopt{\citetads}[3][][]{\href{http://adsabs.harvard.edu/abs/#3}%
    {\def\hyper@linkstart##1##2{}%
     \let\hyper@linkend\@empty\citet[#1][#2]{#3}}}
  \newcommandtwoopt{\citeyearads}[3][][]%
    {\href{http://adsabs.harvard.edu/abs/#3}
    {\def\hyper@linkstart##1##2{}%
     \let\hyper@linkend\@empty\citeyear[#1][#2]{#3}}}
\makeatother

\begin{document}

\titlerunning{Fast sdB stars from the Hyper-MUCHFUSS project}
  \title{Spectroscopic twin to the hypervelocity sdO star US 708 and three fast sdB stars from the Hyper-MUCHFUSS project}

  \author{E.~Ziegerer\inst{\ref{inst1}} \and U.~Heber\inst{\ref{inst1}} \and S.~Geier\inst{\ref{inst1},\ref{inst2},\ref{inst3},\ref{inst4}} \and A.~Irrgang\inst{\ref{inst1}} \and T.~Kupfer\inst{\ref{inst6}} \and F.~F\"urst\inst{\ref{inst6},\ref{inst7}} \and J.~Schaffenroth\inst{\ref{inst1}}}
  \institute{Dr.~ Karl Remeis-Observatory \& ECAP, Astronomical Institute, Friedrich-Alexander University Erlangen-N\"urnberg, Sternwartstr.~7, 96049 Bamberg, Germany \email{Eva.Ziegerer@fau.de}\label{inst1} \and European Southern Observatory, Karl-Schwarzschild-Str.~2, 85748 Garching, Germany \label{inst2} \and Department of Physics, University of Warwick, Coventry CV4  AL, UK \label{inst3} \and Institute for Astronomy and Astrophysics, Kepler Center for Astro and Particle Physics, Eberhard Karls University, Sand 1, 72076 T\"ubingen, Germany \label{inst4} \and Division of Physics, Mathematics, and Astronomy, California Institute of Technology, Passadena, CA 91125, USA \label{inst6} \and European Space Astronomy Centre (ESA/ESAC), Operations Department, Villanueva de la Ca\~nada (Madrid), Spain \label{inst7}}

  \date{Received / Accepted}
  
  \abstract{
Important tracers for the dark matter halo of the Galaxy are hypervelocity stars (HVSs), which are faster than the local escape velocity of the Galaxy and their slower counterparts, the high-velocity stars in the Galactic halo.  
Such HVSs are believed to be ejected from the Galactic Centre (GC) through tidal disruption of a binary by the super-massive black hole (Hills mechanism).  
The Hyper-MUCHFUSS survey aims at finding high-velocity potentially unbound hot subdwarf stars.
We present the spectroscopic and kinematical analyses of a He-sdO as well as three candidates among the sdB stars using optical Keck/ESI and VLT (Xshooter, FORS) spectroscopy. 
Proper motions are determined by combining positions from early-epoch photographic plates with those derived from modern digital sky surveys. 
The Galactic rest frame velocities range from 203 km\,s$^{-1}$ to 660 km\,s$^{-1}$, indicating that most likely all four stars are gravitationally bound to the Galaxy. 
With $T_\text{eff}=47000$ K and a surface gravity of $\log g = 5.7$,  SDSS J205030.39$-$061957.8 (J2050) is a spectroscopic twin of the hypervelocity He-sdO US 708. 
As for the latter, the GC is excluded as a place of origin based
on the kinematic analysis. 
Hence, the Hills mechanism can be excluded for J2050. 
The ejection velocity is much more moderate ($385\pm79$ km\,s$^{-1}$) than that of US 708 ($998\pm68$ km\,s$^{-1}$). 
The binary thermonuclear supernova scenario suggested for US 708 would explain the observed properties of J2050 very well without pushing the model parameters to their extreme limits, as required for US 708. 
Accordingly, the star would be the surviving donor of a type Ia supernova.
Three sdB stars also showed extreme kinematics; one could be a HVS ejected from the GC, whereas the other two could be ejected from the Galactic disk through the binary supernova mechanism.
Alternatively, they might be extreme halo stars.
}
  

  \maketitle
  
  \section{Introduction}
  \label{sec_introduction}
Hypervelocity stars (HVSs) are stars that move so fast that they may exceed the escape velocity of the Galaxy.
In the late 1980s, it was predicted by \citetads{1988Natur.331..687H} from numerical experiments that a star can be ejected from the Galaxy with velocities exceeding the escape velocity by the disruption of a binary through tidal interaction with a super-massive black hole (SMBH).
The first such stars were discovered serendipitously in 2005 (\citeads{2005ApJ...622L..33B}; \citeads{2005A&A...444L..61H}; \citeads{2005ApJ...634L.181E}).
However, \citetads{2007ApJ...671.1708B} showed that about 50\% of the ejected stars undergoing this mechanism remain bound to the Galaxy.
We use the term HVS only for stars that are truly unbound.
Interestingly, the nature, number, and distribution of the so-called S-stars, which are normal main-sequence B-stars in the central arcsecond of the Galaxy on close eccentric orbits around the SMBH, are consistent with expectations for the former companions of HVS (\citeads{2008MNRAS.383L..15S}; \citeads{2014ApJ...784...23M}).

In their survey for unbound stars, \citetads{2014ApJ...787...89B} discovered 21 unbound HVSs and 17 lower velocity stars of spectral type B with masses between 2.5 and 4 $M_\odot$, which means that these stars have short lifetimes.
The Galactic Centre (GC) is the only place in our Galaxy known to host an SMBH (\citeads{2003ApJ...596.1015S}; \citeads{2005ApJ...620..744G}; \citeads{2009ApJ...692.1075G}), and therefore the GC is considered the likely place of origin of HVSs. 

The Hills scenario has been studied in many variations.
This includes binary SMBHs, binaries consisting of an SMBH and an intermediate-mass black hole, triple star disruption, in-spiral of a young stellar cluster forming jets of HVSs, and many other numerical calculations (for details we refer to the review by \citeads{2015ARA&A..53...15B}).
There is evidence for a GC origin for the best-studied HVSs (e.g. \citeads{2012ApJ...754L...2B}).
However, the lack of proper motions or their inaccuracy \citepads{2015ApJ...804...49B} prevents the development of Galactic trajectories for most HVS stars to trace their place of origin.
The Hills scenario was challenged by some brighter HVS B-type stars (e.g. HD271791, \citeads{2008A&A...483L..21H}, \citeads{2008ApJ...684L.103P}; HIP 60350, \citeads{2010ApJ...711..138I}) because the GC could be excluded as a place of origin.
HE 0437--5439 is another particularly interesting case, because its time of flight is far too long for it being ejected as a single star from the GC.
A possible origin in the Large Magellanic Cloud is under debate (\citeads{2005ApJ...634L.181E}; \citeads{2008A&A...480L..37P}; \citeads{2010ApJ...719L..23B}; \citeads{2013A&A...549A.137I}; \citeads{2015ApJ...804...49B}).
\citetads{2009ApJ...697.2096P} suggested that a close hypervelocity binary could be ejected from a hierarchical triple through interactions with the SMBH in the GC.
During their stellar evolution it is possible for such close binaries to evolve to mass transfer configurations, and they may even merge to form a blue straggler, which would be sufficiently long-lived.
 
Another mechanism to accelerate stars is the dynamical ejection from open clusters \citepads{1991AJ....101..562L}.
During a close encounter large kicks can be transferred to the least massive of the involved components.
This process is most efficient when two close binaries collide.
Several hundred km\,s$^{-1}$ can be reached, but only at rates that cannot account for a significant fraction of the observed population of HVSs in the Galactic halo \citepads{2012ApJ...751..133P}.

\citetads{1961BAN....15..265B} first proposed the binary supernova ejection mechanism.
When a massive primary undergoes a core-collapse supernova explosion, its secondary is released with an ejection velocity that is closely connected to the secondary's orbital velocity \citepads{1998A&A...330.1047T}.

\citetads{2009ApJ...691L..63A} predicted that the disruption of satellite galaxies may contribute halo stars by stripping them from their host.
The stars may reach velocities exceeding the escape velocity of the Galaxy.
This scenario would form a cluster of HVSs in the sky.
A large portion of the HVSs from the survey of \citetads{2014ApJ...787...89B} indeed cluster around the constellations of Leo and Sextans.

\section{High-velocity hot subdwarfs}
HVSs were also found among hot subdwarf stars.
Subluminous stars of spectral type B and O (sdB, sdO) are likely formed out of a red giant star (RG) that has lost almost its entire hydrogen envelope.
The remaining layer of hydrogen does not have enough mass to sustain a hydrogen-burning shell, like in cooler horizontal branch stars, and sdO/Bs cannot evolve in the canonical way by ascending the asymptotic giant branch before they finally settle on the white dwarf cooling tracks (see \citeads{2009ARA&A..47..211H}, \citeyearads{2016PASP..128h2001H} for reviews).
How the stars are originally stripped of their hydrogen envelope remains under debate.
Systematic surveys revealed that a large portion (40-70\%) of hot subdwarfs are members of close binaries (\citeads{2001MNRAS.326.1391M}; \citeads{2003MNRAS.338..752M}; \citeads{2011MNRAS.415.1381C}; \citeads{2015A&A...577A..26G}), with mostly white dwarfs or low-mass late-type main-sequence stars as companions.
Substellar companions are also known, however, like brown dwarfs \citepads{2015A&A...576A.123S}.
Wide binaries with F, G, K companions and orbital periods of $\sim$ 1000 d exist and may be formed by stable Roche-lobe overflow (\citeads{2012A&A...548A...6V}, \citeyearads{2013A&A...559A..54V}; \citeads{2013ApJ...771...23B}). 
While the close binaries can be explained by a common envelope and spiral-in phase during the RG phase, single hot subdwarfs are less straightforward to explain through mergers of helium white dwarfs, common-envelope mergers, or internal mixing processes (see \citeads{2016PASP..128h2001H}). 
These scenarios are of particular interest to explain the properties of extremely helium-rich O-type subdwarfs (He-sdO).

The only known unbound subluminous HVS, US 708, is such a He-sdO. 
It was discovered by \citetads{2005A&A...444L..61H} as the second HVS. 
The spectroscopic reobservation of US 708 and ground-based proper motion measurements showed that it is the fastest unbound star known so far \citepads{2015Sci...347.1126G}.
Ground-based as well as Hubble Space Telescope proper motion measurements by \citetads{2015Sci...347.1126G} and \citetads{2015ApJ...804...49B} exclude an origin in the GC and therefore the Hills ejection mechanism. 
Since the Hills scenario is not valid, a binary supernova scenario has been proposed for the ejection of US 708. 
\citetads{2015Sci...347.1126G} suggested that US 708 is most likely the ejected donor remnant of a thermonuclear supernova (SN Ia) after it was spun up by the tidal interaction with its former close white dwarf companion.
Because the orbit shrinks as a result of the radiation of gravitational waves, US 708 started to transfer helium-rich matter to its compact white dwarf companion.
After a critical mass was deposited on the surface of the white dwarf, the helium ignited and triggered the explosion of the C/O core of the white dwarf.
This so-called double detonation has been proposed as the cause for underluminous SN Ia (\citeads{2009ApJ...707L.163S}; \citeads{2011MNRAS.416.2607G}).
\citetads{2013A&A...554A..54G} identified the sdB binary CD--30$^\circ$ 11223 as a progenitor candidate for such a scenario.
The sdB in CD--30$^\circ$ 11223 has been spun up by the tidal influence of the close white dwarf companion to a projected rotational velocity $\varv_\text{rot}\sin i \simeq 180\,\text{km\,s}^{-1}$, which is significantly higher than the rotation that was found for single sdBs ($<10\,\text{km\,s}^{-1}$, \citeads{2012A&A...543A.149G}). 
An ejected remnant is predicted to have similarly high $\varv_\text{rot}\sin i$ \citepads{2013ApJ...773...49P}.

The Hyper-MUCHFUSS project was started with the aim to find potentially unbound hot subdwarfs.
Twelve sdB stars have been found during the first campaign \citepads{2011A&A...527A.137T}.
One of the goals is to distinguish between an old bound population of hot subdwarfs in the Galactic halo and the possibly unbound ejected SN Ia donor remnants similar to US 708.
In the latter case, they are predicted to be fast rotators spun up by the tidal influence of their close companions.
In the former case, they were formed as single stars and are expected to be slow rotators just like the single sdBs in the field \citepads{2012A&A...543A.149G}.
(\citeads{2008A&A...484L..31H}). 

The possibly unbound Hyper-MUCHFUSS sdB SDSS J121150.27$+$143716.2 (short J1211) is of particular interest because we discovered a cool companion to this sdB star \citepads{2016ApJ...821L..13N} orbiting through the outermost parts of the Milky Way.
This immediately excludes the SN channel.
An origin from the GC and the acceleration there through the slingshot mechanism was also excluded.
First, because the binary is too wide to have survived the destruction of a hierarchical triple.
Second, its kinematics in the past do not point to the GC.
\citetads{2016ApJ...821L..13N} suggested the formation in the halo or the accretion from the tidal debris of a dwarf galaxy that was disrupted by the Milky Way \citepads{2009ApJ...691L..63A}.

In the following section we present the spectroscopic and kinematic analysis of four interesting hot subdwarfs.
The sdB star SDSS J123137.56$+$074621.7 (J1231) and the He-sdO star SDSS J205030.39$-$061957.8 (J2050) have been discovered as new objects with extreme kinematics \citepads{2015A&A...577A..26G}. 
SDSS J163213.05$+$205124.0 (J1632) and SDSS J164419.44$+$452326.7
(J1644) have previously been investigated by \citetads{2011A&A...527A.137T} and are now revisited here.
They were reobserved with higher quality data to improve the constraints on their origins and kinematics.

\section{Observations, atmospheric parameters, and spectroscopic distances}

Preliminary atmospheric parameters, spectroscopic distances, and radial velocities have been obtained from low-resolution SDSS spectra for the preselection of interesting candidates.
For the more accurate analyses presented here we used spectra taken with the SDSS/BOSS, Keck/ESI, ESO-VLT/XSHOOTER, and ESO-VLT/FORS1 spectrographs.
The ESI spectra have been reduced with the pipeline Makee\footnote{\url{http://www.astro.caltech.edu/~tb/makee/}}.
Pipeline-reduced BOSS and XSHOOTER spectra have been downloaded from the SDSS and the ESO Phase 3 databases, respectively.
The reduction of the FORS1 spectra is described in \citetads{2011A&A...527A.137T}.
Details about wavelength coverage and resolution of the spectra are provided in Table~\ref{tab_spectra}.

All spectra were used to search for radial velocity variations in order to search for possible companions.
Therefore, we fitted a set of mathematical functions (Gaussians, Lorentzians and polynomials) to the hydrogen Balmer lines, and if present, to helium lines.
The FITSB2 routine by \citetads{2004Ap&SS.291..321N} was applied as well as the spectrum plotting and analysis suite (SPAS) developed by \citetads{2009PhDT.......273H}.
No radial velocity variations were detected within the uncertainties.

\subsection{Atmospheric parameters}
\begin{figure}[t]
\begin{center}
\includegraphics[scale=0.4]{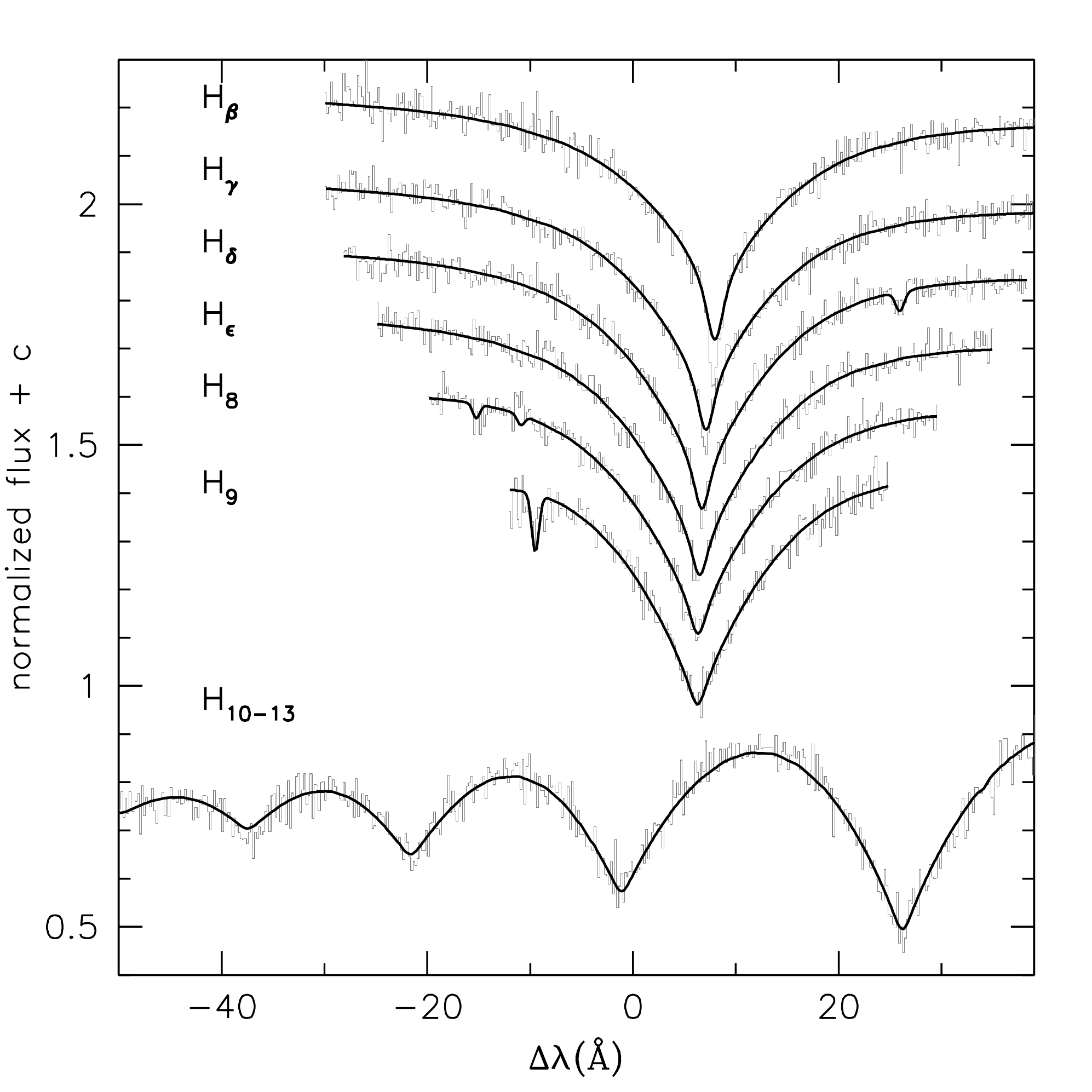}
\caption{\label{fig_J1231_H}Fit of a model spectrum (full line) of the Balmer series for J1231 with the XSHOOTER observation spectrum (grey).}
\end{center}
\end{figure}

\begin{figure}[t]
\begin{center}
\includegraphics[scale=0.4]{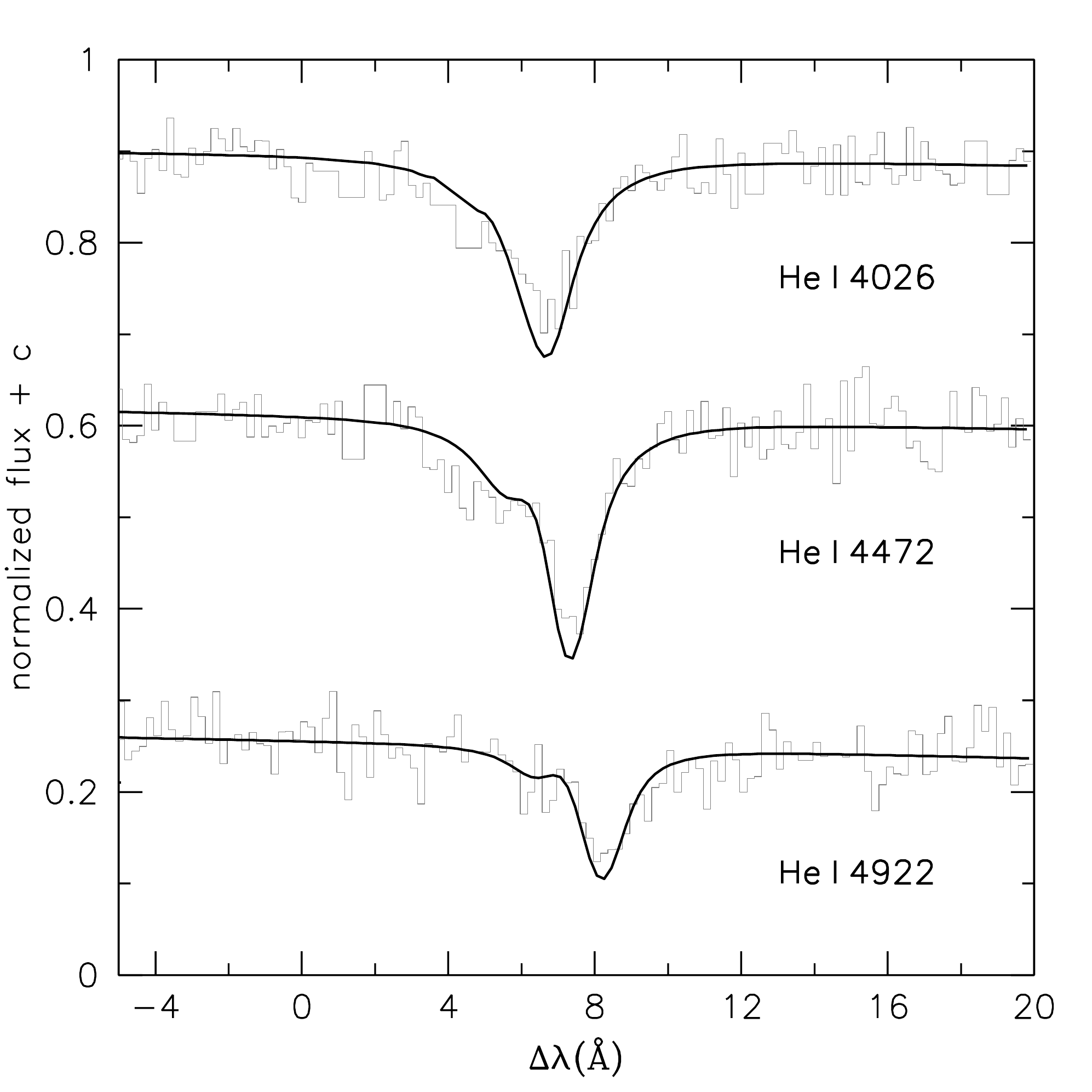}
\caption{\label{fig_J1231_He}Fit of a model spectrum (full line) of HeI lines for J1231 with the XSHOOTER observation spectrum (grey).}
\end{center}
\end{figure}

\begin{figure}[t]
\begin{center}
\includegraphics[scale=0.4]{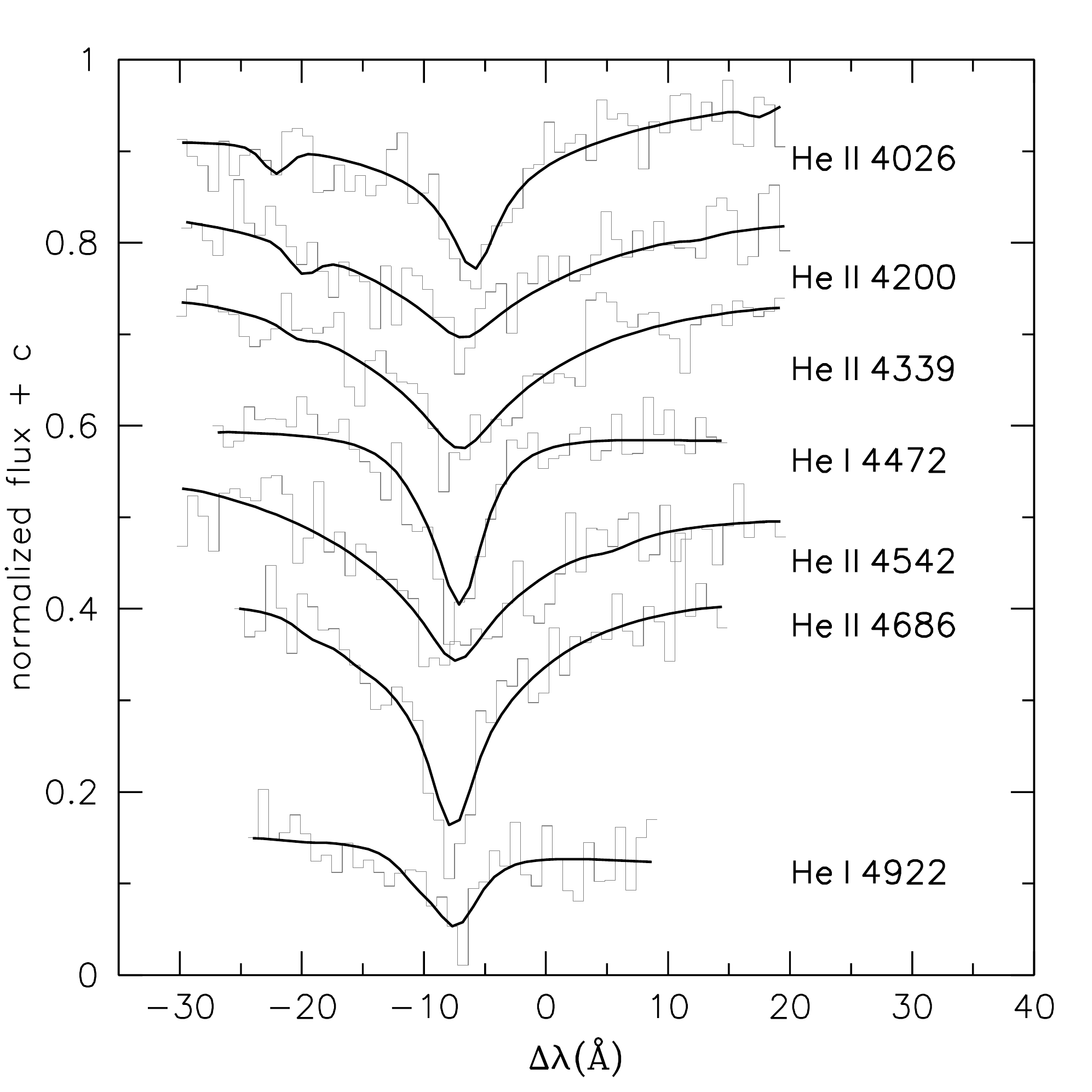}
\caption{\label{fig_J2050}Fit of a model spectrum (full line) of He lines for J2050 with the FORS1 observation spectrum (grey).}
\end{center}
\end{figure}

\begin{figure}[t]
\begin{center}
\includegraphics[scale=0.4]{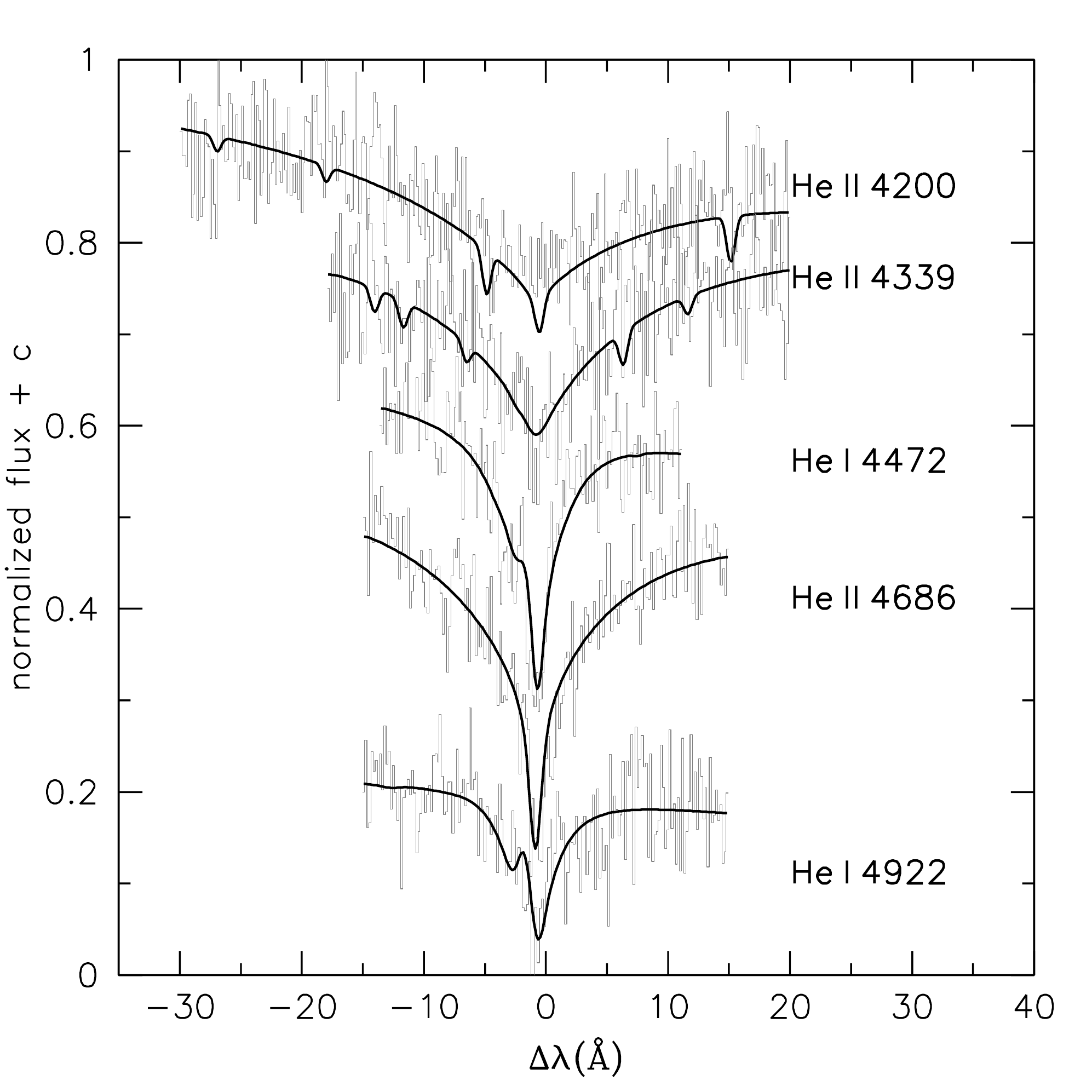}
\caption{\label{fig_J2050_esi}Fit of a model spectrum (full line) of He lines for J2050 with the ESI observation spectrum (grey).}
\end{center}
\end{figure}

A quantitative spectral analysis also provided the atmospheric parameters effective temperature $T_\text{eff}$, surface gravity $\log g$, and helium abundance, as well as limits on the projected rotational velocity $\varv_\text{rot}\sin i$. 
We applied the method described in \citetads{2005A&A...430..223L} and \citetads{2007A&A...462..269S}.
To determine the atmospheric parameters, we fitted the Balmer, He\,{\sc i}, and He\,{\sc ii} lines with model spectra by means of $\chi^2$-minimization using the SPAS routine \citepads{2009PhDT.......273H}.
For the sdB stars with temperatures $T_\text{eff}$ lower than 30000K (J1231 and J1632) we used a grid of metal line-blanketed local thermal equilibrium (LTE) model spectra of \citetads{2000A&A...363..198H} with solar metallicity.
For the one star with $T_\text{eff}$ greater than 30000K  (J1644) we used LTE model spectra with enhanced metal line-blanketing of \citetads{2006A&A...452..579O}.
For the He-sdO star (J2050) we applied the NLTE model spectra  of  \citetads{2009JPhCS.172a2015H} that take into account the line-blanketing caused by nitrogen and carbon.
The adopted uncertainties are typical systematic deviations between different models (see \citeads{2007A&A...464..299G} for details).
The statistical uncertainties based on a bootstrapping algorithm are smaller in all cases.
The results are listed in Table \ref{tab_atmo}.
Figures \ref{fig_J1231_H} and \ref{fig_J1231_He} show the best fit of a model spectrum with the XSHOOTER spectrum of J1231 for the region of the Balmer series and HeI lines, respectively.
As illustrative examples, Fig. \ref{fig_J2050} and \ref{fig_J2050_esi} show the best fit of a model spectrum for the region of HeI and HeII lines with the FORS1 and ESI spectrum of J2050.

The three sdB stars have typical effective temperatures. 
However, it is worth mentioning that the low gravity of J1231 implies that the star is close to termination of core helium burning, possibly even beyond that phase.
The helium content of J1231 and J1632 is typical for the majority of sdB stars.
However, we could not detect any helium lines in the hot J1644, which implies that its abundance ($\text{He}/\text{H}<1/1000$) is considerably lower than expected for sdBs of similar temperature \citepads{2003A&A...400..939E}.
The sdO star J2050 does not show any hydrogen, and we were only
able to derive a lower limit of the helium-to-hydrogen ratio of 100.
Its temperature, gravity, and helium content are typical for He-sdO stars \citepads{2007A&A...462..269S}, in particular similar to that of the hyper-velocity sdO star US 708 \citepads{2015Sci...347.1126G}.
For US 708 an unexpected high projected rotational velocity of $\varv_\text{rot}\sin i=115\pm8$km\,s$^{-1}$ was found.
In comparison to US 708, all program stars show moderate $\varv_\text{rot}\sin i<45$km\,s$^{-1}$.

\subsection{Spectroscopic distances}

From the atmospheric parameters and the apparent visual magnitude we derived the spectroscopic distance as described in \citetads{2001A&A...378..907R}.
For J1231 and J1632 we adopted the atmospheric parameters that were obtained from the XSHOOTER spectra, as they have the highest resolution and a wide wavelength range so that the Balmer jump is accessible, which is very sensitive to the gravity $\log g$.
For the remaining two stars we adopted a mean value of the results from ESI and BOSS (J1644), and ESI and SDSS (J2050) spectra, respectively.

The SDSS g and r magnitudes were converted into Johnson V magnitudes\footnote{\url{http://www.sdss.org/dr6/algorithms/sdssUBVRITransform.html}}, which then were corrected for interstellar reddening.
The reddening was found using a dust extinction tool that gives the Galactic dust reddening for a line of sight\footnote{\url{irsa.ipac.caltech.edu/applications/DUST}}.

\begin{table*}
\caption{\label{tab_atmo} Atmospheric parameters}
\renewcommand{\arraystretch}{1.4}
\setlength{\tabcolsep}{1mm}
\begin{tabular}{lllcccccccc}
\hline\hline
Name & Short & Type & $V$ & $A_V$ & $T_\text{eff}$ & $\log g$ & $\frac{\log n(\text{He})}{\log n(\text{H})}$ & $\varv_\text{rot}\sin i$ & $\varv_\text{rad}$ & $d$ \\
 & & & (mag) & (mag) & (K) & (cgs) & & (km/s) & (km/s) & (kpc) \\
\hline
SDSS J123137.56+074621.7 & J1231 & sdB & 17.44 & 0.05 & $25200\pm500$ & $5.13\pm0.05$ & $-2.23\pm0.05$ & $<45$ & $467\pm2$ & $6.3^{+0.5}_{-0.5}$ \\
SDSS J163213.05+205124.0 & J1632 & sdB & 17.62 & 0.15 & $28900\pm500$ & $5.61\pm0.05$ & $-1.83\pm0.03$ & $<33$ & $-239\pm4$ & $4.3^{+0.3}_{-0.3}$ \\
SDSS J164419.44+452326.7 & J1644 & sdB & 17.39 & 0.03 & $33600\pm500$ & $5.73\pm0.05$ & $<-3.0$ & $<38$ & $-309\pm9$ & $4.1^{+0.3}_{-0.3}$ \\
SDSS J205030.39--061957.8 & J2050 & He-sdO & 18.22 & 0.20 & $47500\pm1000$ & $5.70\pm0.1$ & $>+2.0$ & $<38$ & $-509\pm19$ & $7.0^{+0.9}_{-0.8}$ \\
\hline
\end{tabular}                                                  
\centering
\tablefoot{$V$ is the apparent magnitude, $A_V$ is the reddening in $V$, and $d$ is the heliocentric distance.}
\end{table*}

\subsection{Spectral energy distribution}

To check whether the spectroscopic values are consistent with photometry, we performed a fit of the observed spectral energy distribution (SED). 
Synthetic SEDs are based on the {\sc Atlas12} code \citepads{1996ASPC..108..160K} using an averaged metal abundance from Fig.~6 in \citetads{2013MNRAS.434.1920N} as baseline metallicity. 
While the effective temperature and surface gravity are fixed to their spectroscopic values, we fitted the angular diameter as a distance scaling factor, the color excess $E(B-V)$ as a measure of interstellar extinction (using the description of \citeads{1999PASP..111...63F}), and the scaled average abundance pattern.
The observed SEDs of our four stars were perfectly matched by the synthetic spectra calculated using the atmospheric parameters derived from spectroscopy.
The obtained values for distances and reddening from SED-fitting fit to those obtained by spectroscopy within their uncertainties. 

\subsection{Search for signatures of potential cool companions}

\begin{figure*}
\centering
\includegraphics[width=1\textwidth]{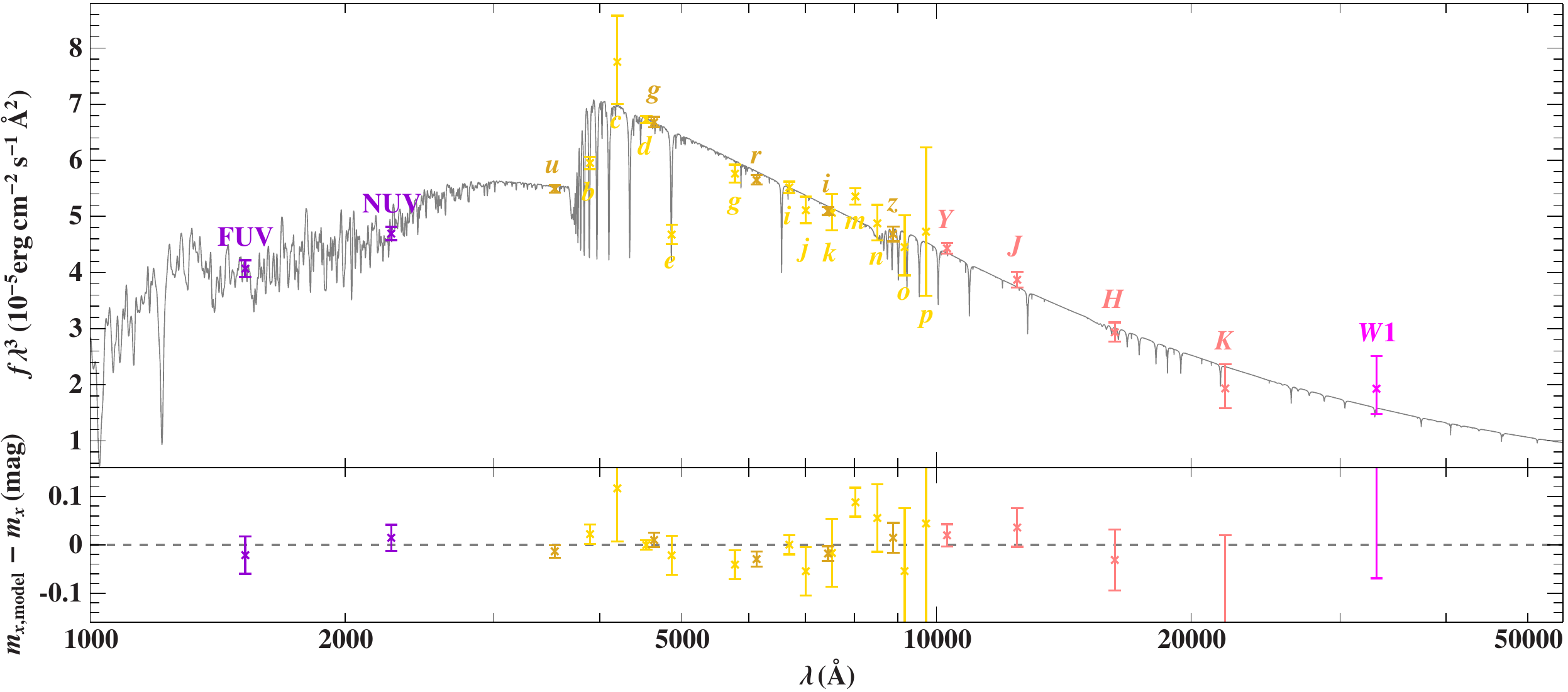}
\caption{Comparison of synthetic and observed photometry of J1231: The \textit{\textup{top panel}} shows the spectral energy distribution. The colored data points are fluxes that are converted from observed magnitudes, and the solid grey line is the model. The residual panel at the \textit{\textup{bottom}} shows the differences between synthetic and observed magnitudes. The photometric systems have the following color code: GALEX (violet), BATC (gold), SDSS (goldenred), UKIDSS (pink), and WISE (magenta).}
\label{fig_SED_J1231}
\end{figure*}

\begin{figure}[t]
\begin{center}
\includegraphics[scale=0.45]{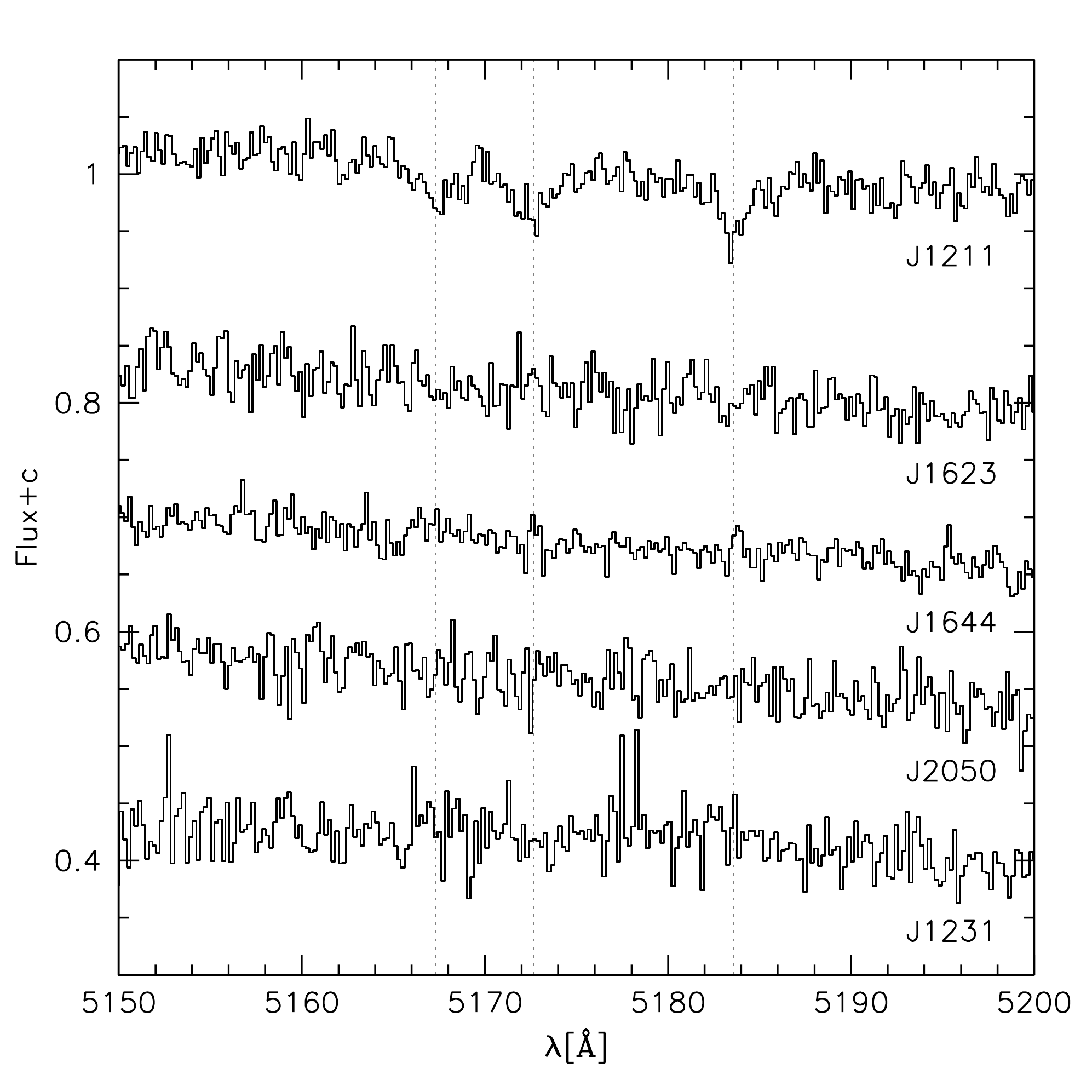}
\caption{\label{fig_Mg_lines}Comparison of the spectra of the four program stars (XSHOOTER for J1231, ESI for J1632, J1644, and J2050) with the spectrum of J1211 in the area of Mg I triplet (marked with dotted vertical lines).}
\end{center}
\end{figure}

Photometric magnitudes from GALEX DR6\footnote{Available in the MAST archive: \url{http://galex.stsci.edu/GR6/?page=mastform}} and SDSS DR12 were available for all stars.
BATC DR1\footnote{\url{http://vizier.cfa.harvard.edu/viz-bin/Cat?II/262}}, UKIDSS DR9 \citepads{2007MNRAS.379.1599L}, and ALLWISE \citepads{2013yCat.2328....0C} were only available for J1231.
Therefore, data in the infrared were only available for one star, and it was possible only for this one star to search for an infrared excess as an indication for a cooler companion (Fig.~\ref{fig_SED_J1231}).
There is no sign of a cooler companion as was seen in the SED of the fast sdB star J1211 \citepads{2016ApJ...821L..13N}.
J1211 was analysed in the same way as for our sample, and we
found a K-type companion that produced an infrared excess in the SED.
Absorption lines of the companion of J1211 were also visible in the spectrum.
No such lines were found in any of our four program stars.
Figure \ref{fig_Mg_lines} shows a comparison of the spectra (XSHOOTER for J1231, ESI for J1632, J1644, and J2050) of the program stars with the spectrum of J1211.
While the spectrum of J1211 shows the Mg\,{\sc i} triplet in the respective area, none is visible in any of the program stars.
Hence, there is no evidence for a cool companion to any of our program stars.

\section{Proper motions}
\label{kap_pm}
The proper motions of the program stars were either taken from \citetads{2011A&A...527A.137T} or determined by the same method as described there.
Early-epoch photographic plates from the Digitised Sky Surveys\footnote{\url{http://archive.stsci.edu/cgi-bin/dss_plate_finder}} were combined with those obtained from the data bases of modern digital surveys such as SDSS\footnote{\url{http://skyserver.sdss3.org/public/en/tools/chart/navi.aspx}}, Super Cosmos\footnote{\url{http://www-wfau.roe.ac.uk/sss/pixel.html
}} , and VHS\footnote{\url{http://www.eso.org/qi/catalogQuery/index/51}}.
This provided a time base of about 60 years.

For each star, positions were derived from all available images with respect to a set of faint, compact, and well-distributed background galaxies.
The galaxies for the reference system are taken from the SDSS database.
We used as many galaxies as possible, but excluded those that show displacements which could be true motion (if the object is misclassified in the SDSS and is in fact a foreground star).
The object was then excluded in all epochs.
It can also be spurious if it is detected only in certain images, which can be caused for instance by a close faint background star that  is detected only in certain wavelength ranges, as the photographic plates are taken in different filters and the companion is only detected in certain filters.
Then these objects are only excluded for those epochs where the motion occurs. 

The comparison of our proper motions with catalogues such as APOP \citepads{2015AJ....150..137Q}, HSOY \citepads{2017arXiv170102629A}, PPMXL \citepads{2010AJ....139.2440R}, SDSS \citepads{2012ApJS..203...21A}, USNO-B1.0 \citepads{2003AJ....125..984M}, and UCAC4 \citepads{2012yCat.1322....0Z} showed that our values are in good agreement within the uncertainties, see Table \ref{tab_pm}.
The HSOY catalogue is a combination of Gaia DR1 and PPMXL data. The resulting values are in good agreement with the values of our proper motion with smaller uncertainties than for PPMXL alone.

For J1644 alone, the values for one of the two proper motion ($\mu_\alpha\cos\delta$) components differ between the different measurements.
Therefore, we discuss two different options for J1644.
First, we use the proper motion obtained by \citetads{2011A&A...527A.137T}, and second, a weighted mean of the catalogue values (denoted as J1644b).
For the remaining program stars we used our proper motion or the one obtained from \citetads{2011A&A...527A.137T} for the further analysis.

\begin{table}
\caption{\label{tab_pm}Proper motions} 
\begin{tabular}{lccl}
\hline\hline
Name & $\mu_\alpha\cos\delta$ & $\mu_\delta$ & catalogue \\
 & (\,mas\,yr$^{-1}$) & (\,mas\,yr$^{-1}$) &  \\
\hline
J1231 & $-7.9\pm3.4$ & $-5.0\pm2.8$ & this paper  \\
 & $-2.5\pm1.3$ & $-6.8\pm1.9$ & APOP \\
 & $-10.5\pm2.5$ & $-7.0\pm2.5$ & HSOY \\
 & $-7.5\pm5.6$ & $-4.1\pm5.6$ & PPMXL  \\
 & $-4\pm3$ & $-2\pm3$ & SDSS \\
J1632 & $-12.5\pm3.0$ & $-1.6\pm3.6$ & T11  \\
 & $-8\pm2.7$ & $-3.5\pm3.4$ & APOP \\
 & $-16.6\pm5.3$ & $-5.8\pm5.3$ & PPMXL \\
 & $-13\pm3$ & $-4\pm3$ & SDSS \\
 & $-10\pm2$ & $0\pm2$ & USNO-B1.0 \\
J1644 & $4.7\pm2.8$ & $-26.1\pm3.3$ & T11  \\
 & $-1.1\pm3.2$ & $-16.4\pm3.2$ & APOP  \\
 & $-7\pm5.9$ & $-27.5\pm5.9$ & PPMXL \\
 & $-1\pm3$ & $-26\pm3$ & SDSS \\
 & $-2\pm6$ & $-26\pm3$ & USNO-B1.0 \\
J2050 & $5.5\pm4.8$ & $-8.9\pm3.5$ & this paper  \\
 & $3.2\pm3.7$ & $-2.9\pm2.1$ & APOP \\
 & $1.8\pm2.4$ & $-9.7\pm2.4$ & HSOY \\
 & $-3.8\pm6.2$ & $-7.5\pm6.2$ & PPMXL \\
 & $0\pm3$ & $-4\pm3$ & SDSS \\
\hline
\end{tabular}                                                  
\centering
\tablefoot{T11: \citetads{2011A&A...527A.137T}}
\end{table}

\section{Kinematics: Extreme halo or ejected stars}

We calculate trajectories of the program stars in three different Milky Way mass models of \citetads{2013A&A...549A.137I} to trace the orbits back to the Galactic disk to obtain their dynamical properties and possible origins.
The halo mass of these three models ranges from $M_{R<200\,\text{kpc}}=1.2-3.0\cdot10^{12}M_\odot$, which covers the whole range of halo masses of other widely used halo mass distributions.
Nevertheless, we tested a fourth mass model \citepads{2017MNRAS.tmp..102R}.
All mass models share the same disk structure \citepads{1975PASJ...27..533M}.
While \citetads{2013A&A...549A.137I} also used their bulge model, \citetads{2017MNRAS.tmp..102R} used the \citetads{1990ApJ...356..359H} model.
Model III of \citetads{2013A&A...549A.137I} and \citetads{2017MNRAS.tmp..102R} used the same potential form for the halo, namely the one suggested by \citetads{1997ApJ...490..493N}. 
However, we recall that the mass model of \citetads{2017MNRAS.tmp..102R} was calibrated to different observational constraints than the mass models of \citetads{2013A&A...549A.137I}, which leads to different halo masses of the two mass models.

Long-term orbits were calculated for 5000 Myr to characterise them in the context of population synthesis.
In order to constrain the place of origin, that is, to determine whether the star was ejected from the Galactic disk or centre, we traced the trajectories back to their last disk crossings and calculated the times of flight and ejection velocities for all mass models.
Through a Monte Carlo simulation of a Gaussian distribution with a depth of $10^6$ , we determined all kinematic parameters of the current location of the stars as well as the values at the time and position of their last disk passage, such as velocity components in Cartesian coordinates ($\varv_x$, $\varv_y$, $\varv_z$), with the Sun lying on the negative $x$-axis and the north Galactic pole being on the positive $z$-axis.
Cylindrical coordinates ($\varv_r$, $\varv_\phi$, $\varv_z$), Galactic rest-frame velocity $\varv_\text{grf}$, and ejection velocity $\varv_\text{ej}$ corrected for the Galactic rotation were also calculated for each of the four mass models.
The input parameters for the simulation are the radial velocity $\varv_\text{rad}$, proper motions ($\mu_\alpha\cos\delta$ and $\mu_\delta$), and spectroscopic distance $d$ with their corresponding uncertainties.
For all program stars the resulting disk passage is independent of the choice of the applied mass model.

From the long-term calculations the $z$-component of the angular momentum $J_z$ as well as the eccentricity $e$ of the orbit are determined.
All resulting velocity components and the probability of being bound for models I, II, and III of \citetads{2013A&A...549A.137I} and the model of \citetads{2017MNRAS.tmp..102R} can be found in Table \ref{tab_kine} ($1\sigma$ uncertainties are given).
As can be seen, the choice of model potential is of no importance because all velocities derived from the different models agree within their mutual uncertainties.
While J1632 is certainly bound to the Galaxy, the probability that J1231, J1644, and J2050 are unbound is also low, regardless of the choice of Galactic potential.
Therefore we conclude that our program stars belong to an old Galactic stellar population and investigate their kinematical properties from long-term evolution of their Galactic orbits.

\begin{figure}[t]
\begin{center}
\includegraphics[scale=0.55]{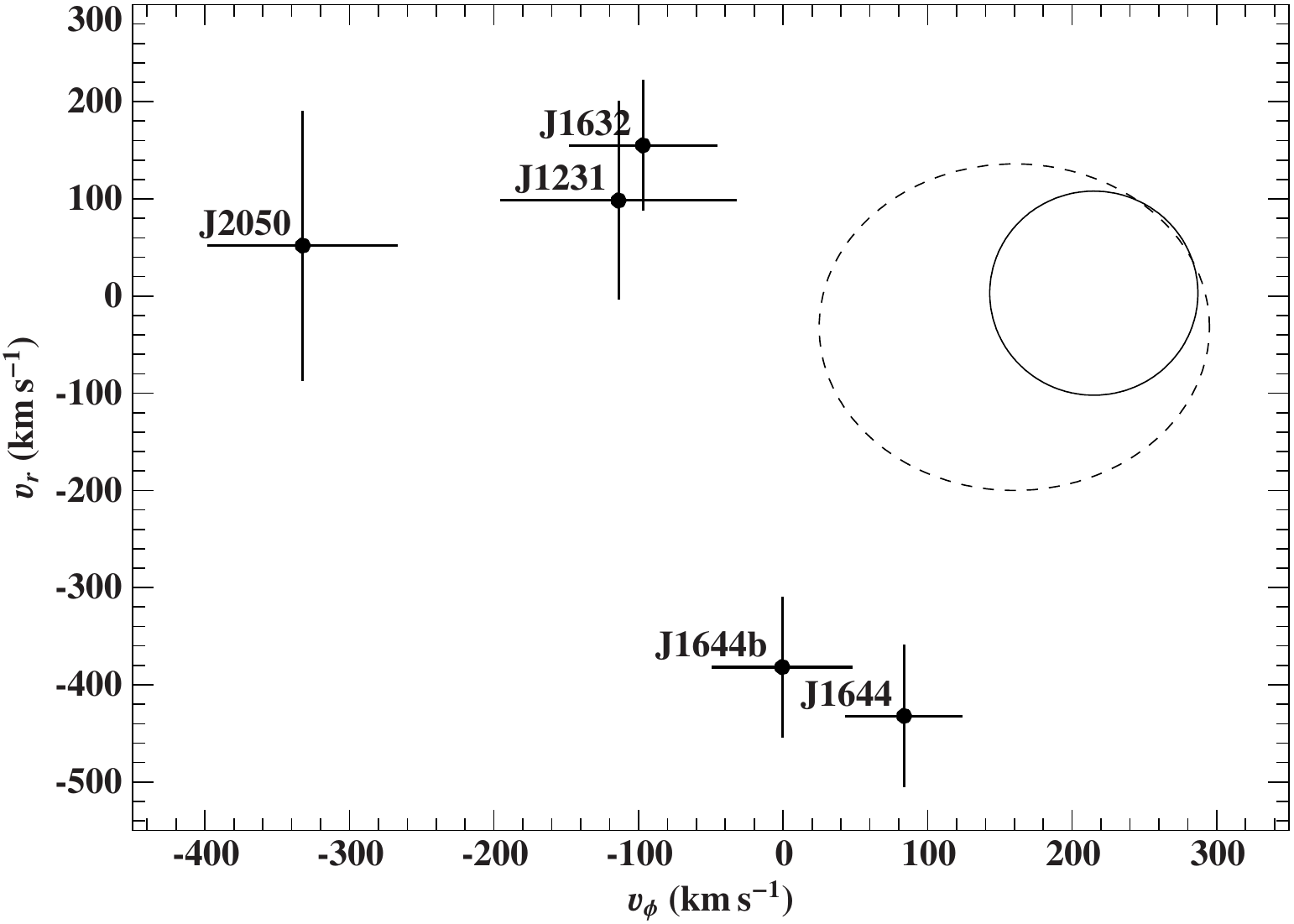}
\caption{\label{fig_UV}Comparison of the program stars in the $\varv_r-\varv_\phi$-diagram with $3\sigma$ contour of the thick disk (dashed line) and $3\sigma$ contour of the thin disk (solid line) according to \citetads{2006A&A...447..173P}.}
\end{center}
\end{figure}

According to \citetads{2006A&A...447..173P}, stars can be assigned to the populations of the different components of the Milky Way - thin disk, thick disk, halo - using three different criteria.
The first is the classification by their position in the $\varv_r-\varv_\phi$-diagram (Fig. \ref{fig_UV}), where $\varv_r$ is the Galactic radial component, which is negative towards the GC, while $\varv_\phi$ is the Galactic rotational component.
Stars that are revolving on retrograde orbits around the GC have negative $\varv_\phi$.
Disk stars are located in a well-defined region.
Thin and thick disk overlap.
Stars that are outside this region are assumed to belong to the Galactic halo.
Figure~\ref{fig_UV} shows the position of the program stars in the $\varv_r-\varv_\phi$-diagram compared to $3\sigma$ contours of the thick and thin disk as introduced by \citetads{2006A&A...447..173P}.
All stars lie well outside the disk region and can therefore be considered as halo stars.

\begin{figure}[t]
\begin{center}
\includegraphics[scale=0.55]{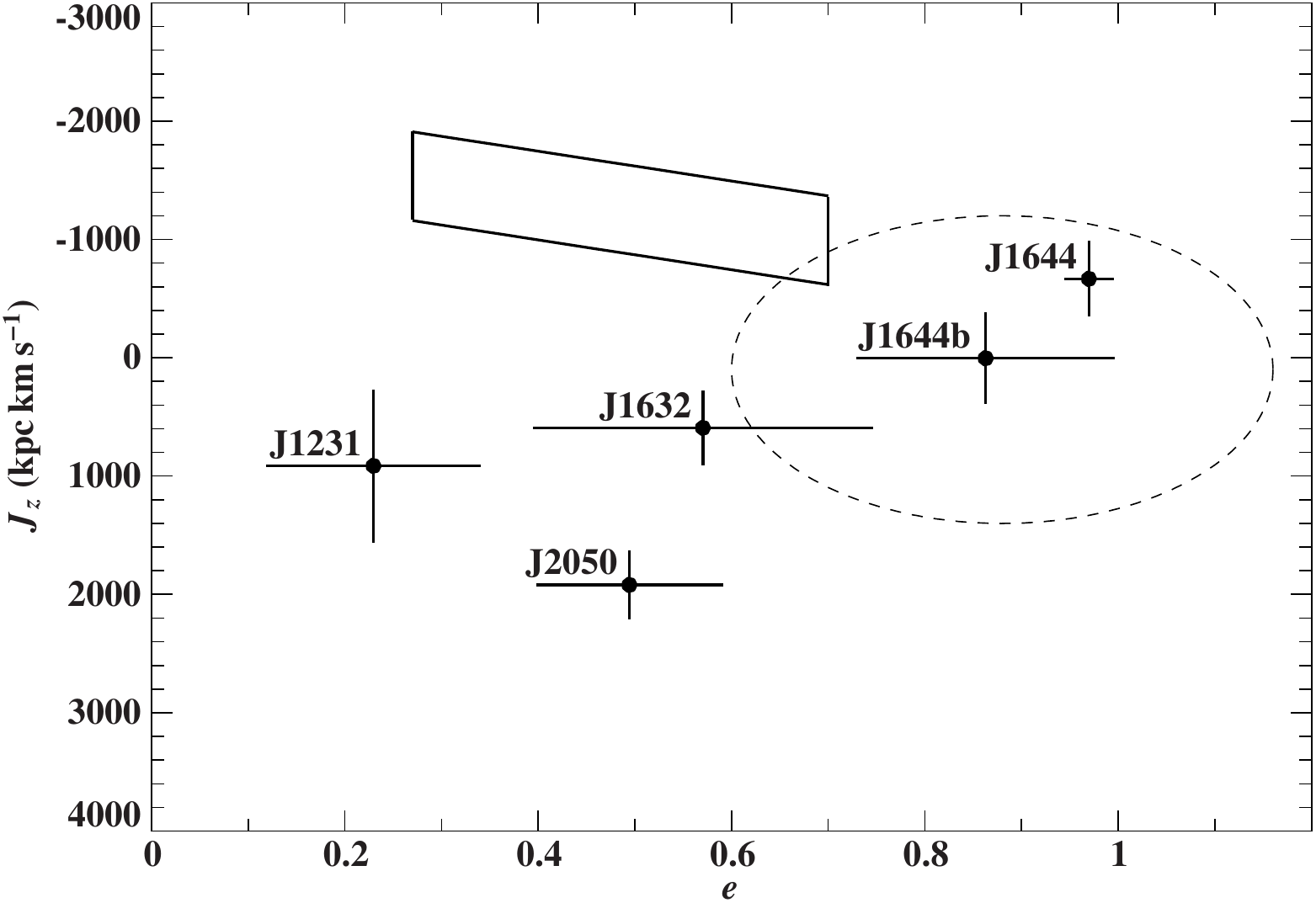}
\caption{\label{fig_Jze}$J_z-e$-diagram, the dashed line indicates the region of typical halo stars. The solid line marks the thick-disk region. Thin-disk stars would populate the continuation of the parallelogram to lower eccentricities.}
\end{center}
\end{figure}

The second diagnostic tool is the $J_z-e$-diagram, which is shown in Fig.~\ref{fig_Jze}.
Stars on retrograde orbits have positive $J_z$.
Thin-disk stars are located at the top left end of the diagram, having very low eccentricities $e$.
\citetads{2011A&A...527A.137T} suggested that stars inside the box belong to the thick disk, while stars inside the ellipse are typical halo stars as they show only little effect of the disk rotation and cross the Galactic plane almost perpendicular on highly eccentric orbits. 
Again, our stars lie well outside the disk region.

The third classification criterion is the shape of the orbit in the $r-z$-diagram itself, where $r$ is the distance of the star to the GC projected onto the Galactic plane $r=\sqrt{x^2+y^2}$.
Thin-disk orbits only cover a very narrow region in this diagram
because they are on very low-eccentricity orbits with very low inclination.
They vary in $r$ by less than 3 kpc and in $z$ by less than 1-2 kpc.
Thick-disk stars show a larger spread in both variables.
Halo objects can have any chaotic orbit imaginable.
The orbits of our stars are discussed individually in the following sections.
For each star the average orbit from the Monte Carlo simulation was calculated 5000 Myr into the past.

\begin{table*}
\caption{\label{tab_kine}Velocity components}
\begin{tabular}{lcccccccc}
\hline
Name/ & $\varv_x$ & $\varv_y$ & $\varv_z$ & $\varv_r$ & $\varv_\phi$ & $\varv_\text{GRF}$ & $\varv_\text{ej}$ & bound\\
Model & (km/s) & (km/s) & (km/s) & (km/s) & (km/s) & (km/s) & (km/s) & \\
\hline
J1231 & $-66\pm97$ & $-136\pm86$ & $378\pm29$ & $99\pm102$ & $-114\pm81$ & $428\pm32$ &\\
I & $-157\pm105$ & $-144\pm76$ & $429\pm27$ & $128\pm102$ & $-152\pm111$ & $495\pm51$& $611\pm67$ & 99.9\% \\
II & $-156\pm105$ & $-146\pm76$ & $429\pm27$ & $127\pm102$ & $-154\pm111$ & $495\pm51$& $612\pm67$ & 99.3\% \\
III & $-154\pm104$ & $-146\pm76$ & $429\pm27$ & $125\pm102$ & $-154\pm110$ & $494\pm51$& $612\pm67$ & 100\%  \\
R & $-153\pm106$ & $-147\pm76$ & $435\pm27$ & $124\pm102$ & $-155\pm111$ & $499\pm51$& $617\pm67$ & 99.5\%  \\
\hline
J1632 & $-179\pm58$ & $-39\pm61$ & $33\pm50$ & $155\pm67$ & $-97\pm51$ & $203\pm54$ & \\
I & $-305\pm58$ & $144\pm79$ & $256\pm63$ & $-4\pm143$ & $-304\pm103$ & $435\pm62$& $612\pm65$ & 100\% \\
II & $-306\pm60$ & $152\pm82$ & $254\pm64$ & $-11\pm145$ & $-309\pm104$ & $438\pm63$& $616\pm65$ & 100\% \\
III & $-302\pm58$ & $131\pm76$ & $263\pm64$ & $16\pm134$ & $-300\pm96$ & $432\pm65$& $609\pm65$ & 100\% \\
R & $-317\pm70$ & $185\pm113$ & $281\pm90$ & $39\pm153$ & $-339\pm114$ & $481\pm92$& $661\pm90$ & 100\%  \\
\hline
J1644 & $433\pm72$ & $-83\pm41$ & $-257\pm42$ & $-432\pm73$ & $84\pm41$ & $514\pm69$ &\\
I & $-467\pm117$ & $-66\pm104$ & $430\pm82$ & $40\pm252$ & $402\pm145$ & $660\pm51$& $553\pm50$ & 91.5\%  \\
II & $-451\pm120$ & $-72\pm112$ & $430\pm88$ & $72\pm250$ & $381\pm149$ & $652\pm52$& $551\pm53$ & 79.3\% \\
III & $-489\pm118$ & $-40\pm95$ & $435\pm78$ & $-1\pm267$ & $410\pm155$ & $675\pm58$& $568\pm55$ & 100\% \\
R & $-489\pm90$ & $-48\pm128$ & $372\pm102$ & $338\pm181$ & $304\pm162$ & $640\pm64$& $572\pm44$ & 84.5\%  \\
\hline
J1644b & $355\pm70$ & $-142\pm51$ & $-162\pm55$ & $-382\pm72$ & $0\pm49$ & $422\pm69$ & \\
I & $-408\pm148$ & $116\pm124$ & $247\pm159$ & $312\pm252$ & $97\pm218$ & $527\pm161$& $546\pm116$ & 99.5\% \\
II & $-396\pm153$ & $114\pm125$ & $237\pm159$ & $316\pm239$ & $87\pm209$ & $511\pm168$& $537\pm120$ & 97.9\% \\
III & $-424\pm136$ & $132\pm119$ & $267\pm159$ & $302\pm277$ & $90\pm231$ & $551\pm149$& $570\pm112$ & 100\% \\
R & $-488\pm90$ & $-48\pm128$ & $372\pm102$ & $338\pm181$ & $304\pm162$ & $640\pm64$& $572\pm44$ & 84.5\%  \\
\hline
J2050 & $-299\pm73$ & $-191\pm71$ & $107\pm92$ & $52\pm139$ & $-332\pm66$ & $394\pm40$ & \\
I & $-114\pm149$ & $46\pm74$ & $-104\pm35$ & $-94\pm144$ & $-107\pm41$ & $215\pm92$& $385\pm79$ & 99.8\% \\
II & $-120\pm146$ & $41\pm74$ & $-102\pm35$ & $-101\pm142$ & $-104\pm41$ & $215\pm93$& $383\pm80$ & 99.6\% \\
III & $-91\pm160$ & $53\pm79$ & $-107\pm38$ & $-69\pm157$ & $-110\pm39$ & $217\pm92$& $389\pm78$ & 100\% \\
R & $-140\pm140$ & $33\pm76$ & $-112\pm35$ & $-118\pm138$ & $-106\pm43$ & $226\pm94$& $390\pm82$ & 99.7\%  \\
\hline
\end{tabular}                                                  
\centering
\tablefoot{Velocity components and the probability of being bound to the Galaxy of the program stars. The values of the first line are the current values, next lines are the values at the last disk passage based on models I, II, and III of \citetads{2013A&A...549A.137I} and the model of \citetads[][R]{2017MNRAS.tmp..102R}, respectively.}
\end{table*}

\subsection{J1644 - an extreme halo star}
The fastest of the program stars is also the most precarious.
Because of the discrepancy in proper motions (see Sect.~\ref{kap_pm}), we carried out the kinematic analyses twice, adopting the proper motions of \citetads{2011A&A...527A.137T} and a weighted mean of the catalogue values, respectively.

\begin{figure*}
\centering
\includegraphics[width=1\textwidth]{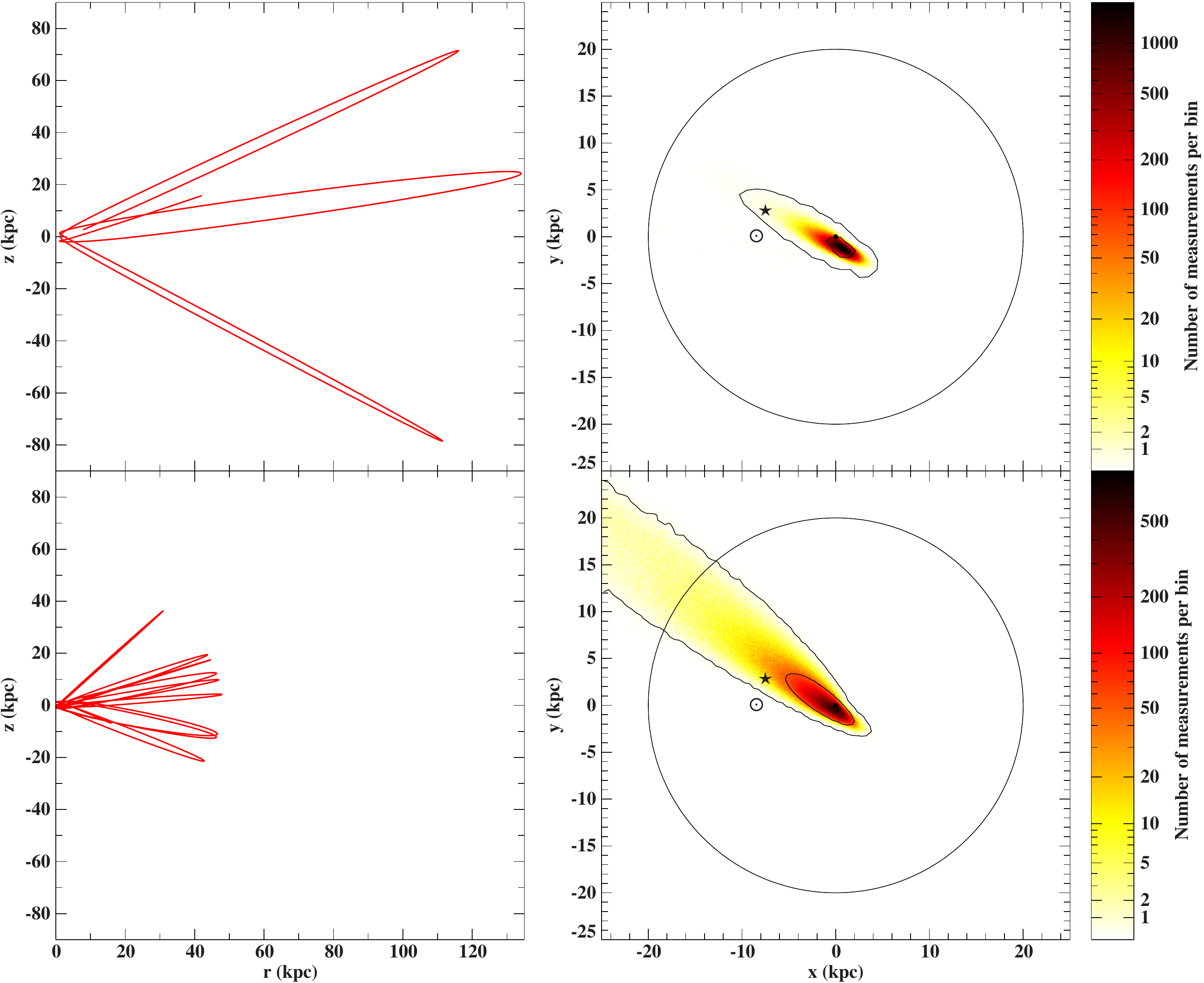}
\caption{\textit{Left panel}: $r-z$-diagram of J1644 using the \citetads{2011A&A...527A.137T} proper motion (top) and the weighted mean of the catalogue proper motions (bottom), respectively. \textit{Right panel:} disk passages binned and colour-coded of J1644 with 1 and 3$\sigma$ contours using the \citetads{2011A&A...527A.137T} proper motion (top) and the weighted mean of the catalogue proper motions (bottom), respectively. The black dot marks the GC, the star the current position of J1644, and the solar symbol the position of the Sun. The circle indicates the Galactic disk. All calculations were performed with model I of \citetads{2013A&A...549A.137I}.}
\label{fig_kine_J1644}
\end{figure*}

Regardless of the choice of the proper motion, J1644 has an extreme kinematic behaviour, as becomes obvious from its position in the $\varv_r-\varv_\phi$- and the $J_z-e$-diagrams (see Figs.~\ref{fig_UV} and~\ref{fig_Jze}).
However, the orbit strongly depends on the choice of their values, as demonstrated in the left panel of Fig.~\ref{fig_kine_J1644}.
When we adopt the proper motion of \citetads{2011A&A...527A.137T}, J1644 is on a highly eccentric orbit, which leads the star to distances of up to $129\pm73$ kpc away from the GC. 
The travel time since its last approach to the GC is much longer, $1558\pm988$ Myr, than the lifetime of an EHB star.
If J1644 were ejected from the GC, it would have been a main-sequence star or subgiant at the time and had to evolve into an sdB on the way.
Adopting the weighted mean of the catalogue proper motion values leads to a shorter travel time of only $113\pm72$ Myr and reaches only distances of $20\pm6$ kpc away from the GC, which is consistent with the EHB lifetime of < 100 Myr, meaning that it is possible to reach the star's current position within the lifetime.
The right-hand panel of Fig.~\ref{fig_kine_J1644} shows that the GC lies within the $1\sigma$ contours of the disk passages,
regardless of the choice of the proper motion.
Although J1644 has the highest $\varv_\text{grf}=514\pm69$ km s$^{-1}$ (or $\varv_\text{grf}=422\pm69$ km s$^{-1}$ for the weighted mean of the catalogue proper motions) of all program stars, it is heading towards the GC and must therefore still be bound.

\subsection{Possibly ejected stars}
\citetads{2011A&A...527A.137T} suggested that halo stars outside the ellipse in the $J_z-e$-diagram (Fig.~\ref{fig_Jze}) could be ejected stars.
Accordingly, J1231, J1632, and J2050, all on retrograde orbits (see Fig.~\ref{fig_UV}), could be runaway stars from the Galactic disk or bound HVS from the GC rather than extreme halo stars.
While J1231, and J2050 cannot originate from the GC (see Figs.~\ref{fig_kine_J1231} and~\ref{fig_kine_J2050}, right panels), the disk-crossing area of the trajectories of J1632 include the GC (see Fig.~\ref{fig_kine_J1632} right panel).
We discuss this object first before addressing the disk runaways J1231 and J2050.

\subsubsection{J1632 - a potentially bound HVS}
\begin{figure*}
\centering
\includegraphics[width=1\textwidth]{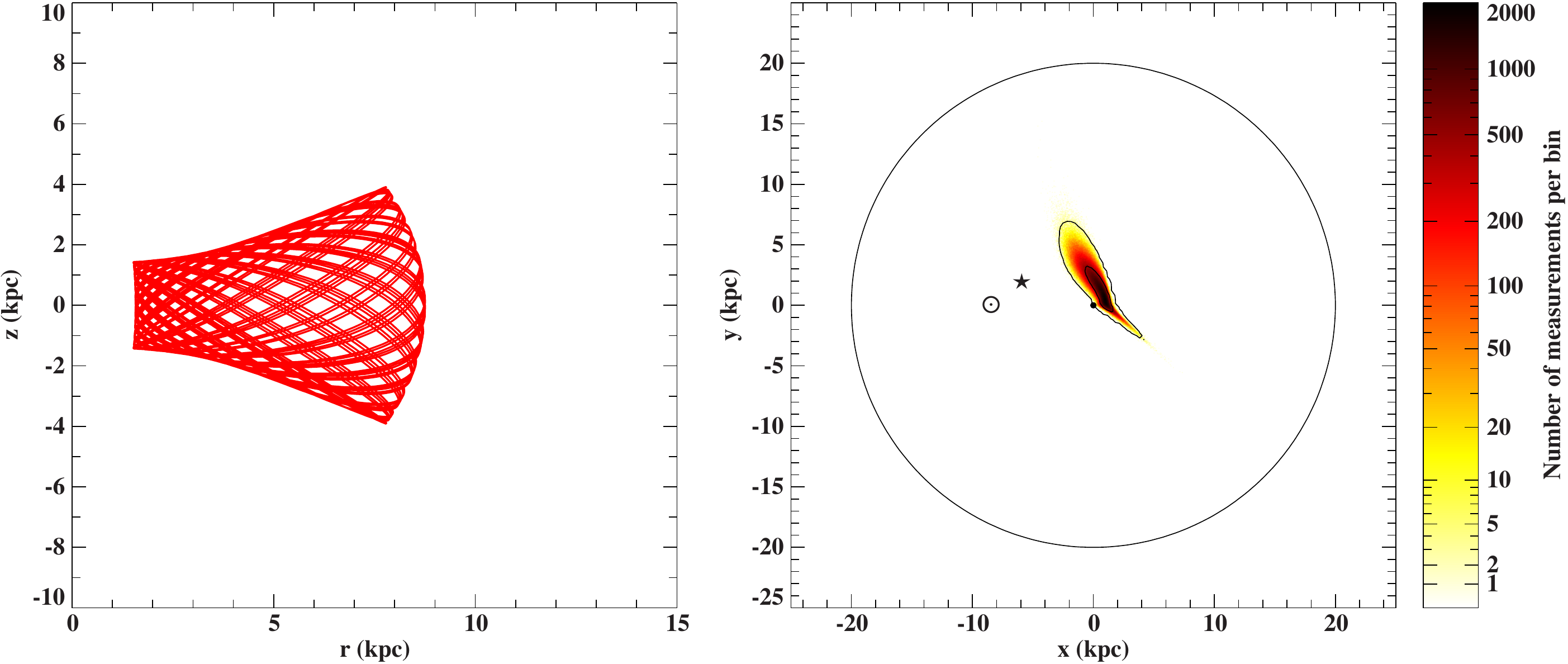}
\caption{\textit{Left panel}: $r-z$-diagram of J1632. \textit{Right panel:} disk passages binned and colour-coded of J1632 with 1 and 3$\sigma$ contours. The black dot marks the GC, the star the current position of J1632, and the solar symbol the position of the Sun. The circle indicates the Galactic disk. All calculations were performed with model I of \citetads{2013A&A...549A.137I}.}
\label{fig_kine_J1632}
\end{figure*}

The analysis of its trajectory indicates that J1632 may originate from the GC and therefore could be a bound HVS (see Fig.~\ref{fig_kine_J1632} left panel).
J1632 has a relatively low $\varv_\text{GRF}=203\pm54\,\text{km\,s}^{-1}$ of the order of typical disk stars and is approaching us.
Similarly, the velocity perpendicular to the Galactic disk is very low ($\varv_z=33\pm50\text{km\,s}^{-1}$), similar to that of a thick-disk star. 
In addition, the eccentricity speaks for a thick-disk star.
Therefore the orbit looks like that of a typical thick-disk star (see left-hand panel of Fig.~\ref{fig_kine_J1632}). 
However, the star is revolving retrograde around the GC, and consequently, it cannot be an ordinary thick-disk star.
\citetads{2015A&A...576A..65R} found an intermediate He-sdB on a similar orbit.
An origin from the GC for J1632 is conceivable (see right-hand panel of Fig.~\ref{fig_kine_J1632}).
Possibly, J1632 could have been ejected into a low-inclination orbit when the former binary was disrupted by the SMBH.
With a travel time of $23.7\pm5.4$ Myr from the GC to its current position, this scenario is consistent with the EHB lifetime of such stars of < 100 Myr.

\subsubsection{J1231 and J2050 - potential disk runaways}
\begin{figure*}
\centering
\includegraphics[width=1\textwidth]{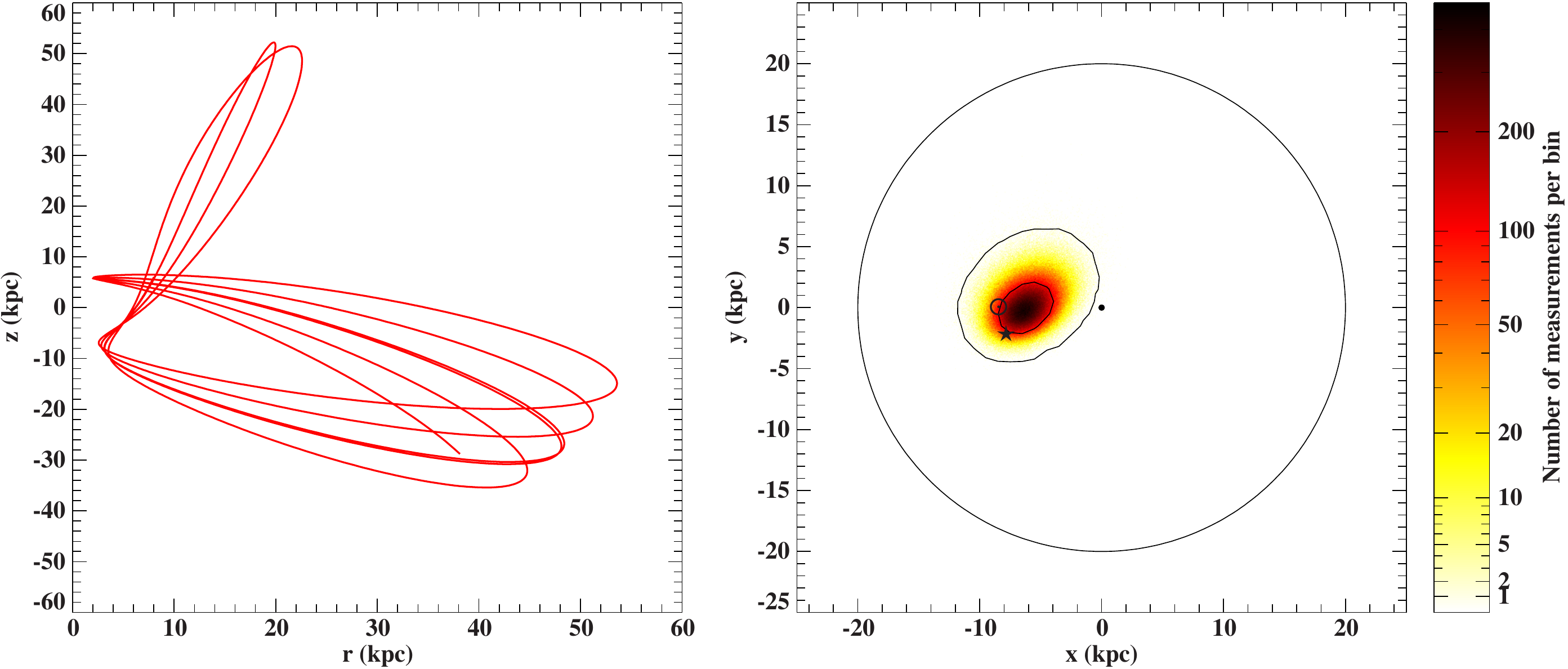}
\caption{\textit{Left panel}: $r-z$-diagram of J1231. \textit{Right panel:} disk passages binned and colour-coded of J1231 with 1 and 3$\sigma$ contours. The black dot marks the GC, the star the current position of J1231, and the solar symbol the position of the Sun. The circle indicates the Galactic disk. All calculations were performed with model I of \citetads{2013A&A...549A.137I}.}
\label{fig_kine_J1231}
\end{figure*}

The constant radial velocity, proper motion, and spectroscopic distance of J1231 indicate a Galactic rest-frame velocity of $v_\text{grf}=428\pm32\,\text{km\,s}^{-1}$ and a likely origin in the Galactic disk rather than the GC (see right-hand panel of Fig.~\ref{fig_kine_J1231}).
A travel time of $14.4\pm1.6$ Myr from the Galactic disk to its current position is consistent with the EHB lifetime of such stars of < 100 Myr.
Although the star is the only one of the program stars receding from us, it is bound with a probability of 99.9\%.
In the context of population membership, J1231 shows a quite chaotic orbit like that of an extreme halo star (see left-hand panel of Fig.~\ref{fig_kine_J1231}). 
It reaches distances of more than 50 kpc away from the Galactic disk.

\begin{figure*}
\centering
\includegraphics[width=1\textwidth]{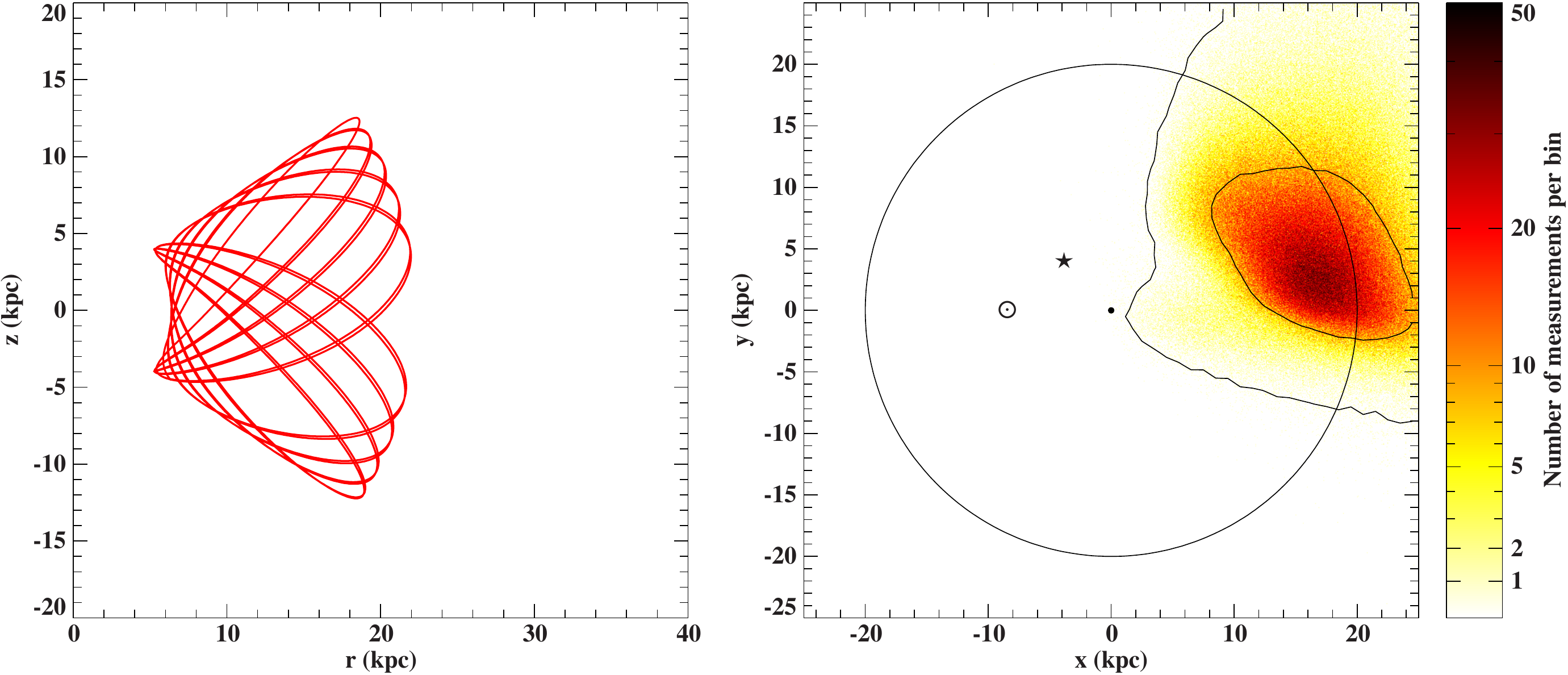}
\caption{\textit{Left panel}: $r-z$-diagram of J2050. \textit{Right panel:} disk passages binned and colour-coded of J2050 with 1 and 3$\sigma$ contours. The black dot marks the GC, the star the current position of J2050, and the solar symbol the position of the Sun. The circle indicates the Galactic disk. All calculations were performed with model I of \citetads{2013A&A...549A.137I}.}
\label{fig_kine_J2050}
\end{figure*}

The traced orbits of J2050 show that the star does not approach anywhere near the GC (see right-hand panel of Fig.~\ref{fig_kine_J2050}) with a typical halo orbit (see left-hand panel Fig.~\ref{fig_kine_J2050}).
With $v_\text{grf}=394\pm40\,\text{km\,s}^{-1}$ on a retrograde orbit, it has a probability of being bound of 99.8\%.
Its travel time from the outskirts of the Galactic disk is $113\pm72$ Myr, which is consistent with the lifetime of EHB stars.

\section{Conclusions}
We have performed a spectroscopic and kinematic follow-up analysis of two known hot subdwarfs from the first Hyper-MUCHFUSS campaign as well as two new ones with extreme kinematics.
Radial velocity measurements, spectral identification, and photometry (when available) were used to exclude binarity or variability of the stars.
Proper motions were either taken from \citetads{2011A&A...527A.137T} or measured in the same way.
The goal of this work was to place constraints on the place of origin of the stars and the possible mechanisms that led the stars to their extreme kinematics.

\begin{figure}[t]
\begin{center}
\includegraphics[scale=0.7]{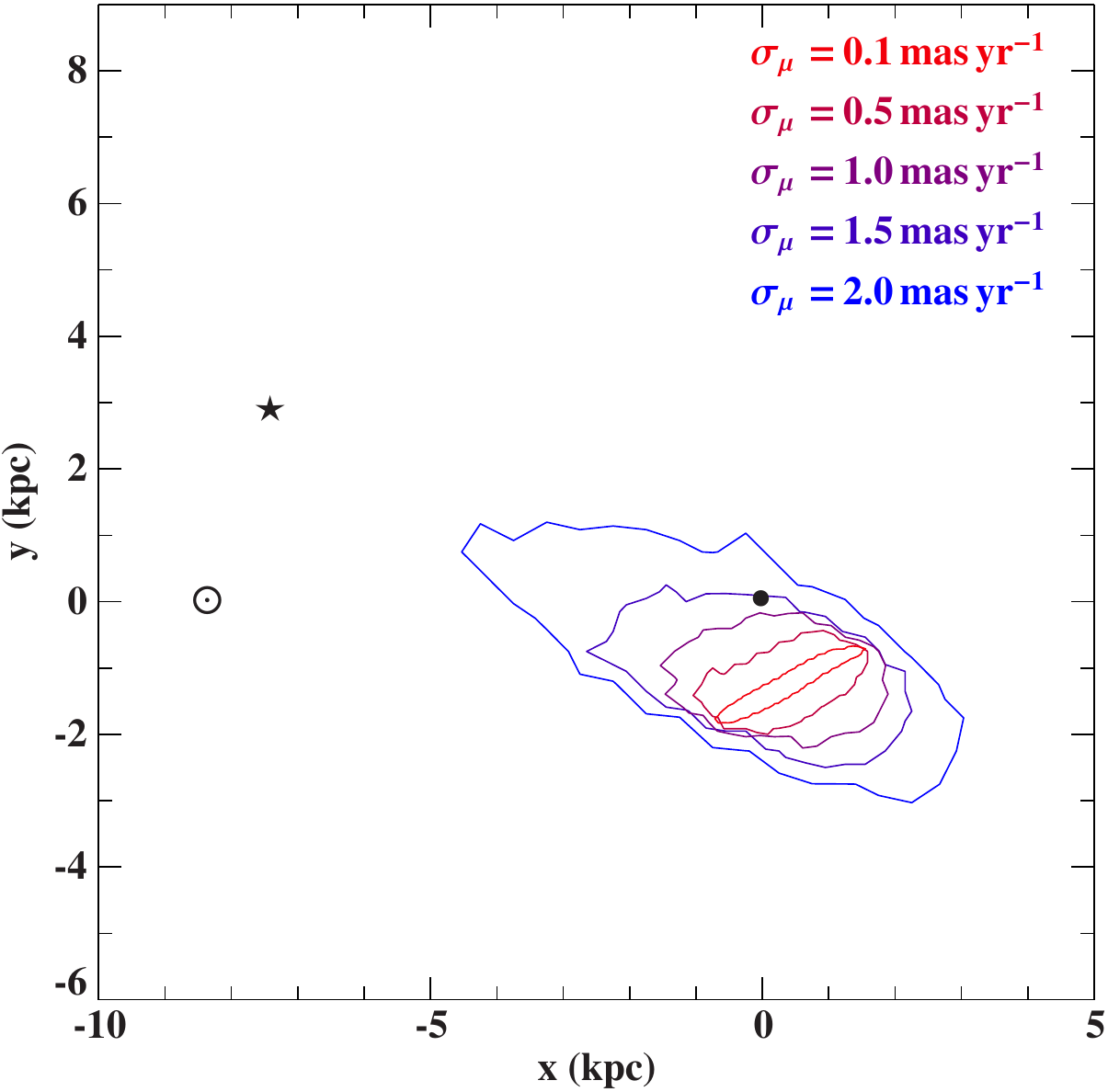}
\caption{\label{fig_kine_J1644_Gaia}Disk passage 3$\sigma$ contour of J1644 using the \citetads{2011A&A...527A.137T} proper motion as in Fig.~\ref{fig_kine_J1644} (upper right) and for uncertainties $\sigma_\mu$ reduced from 2.0\,mas\,yr$^{-1}$ to 0.1\,mas\,yr$^{-1}$. The latter is expected to be the Gaia end-of-mission accuracy. The black dot marks the GC, the star the current position of J1644, and the solar symbol the position of the Sun.}
\end{center}
\end{figure}

While we cannot rule out that the program stars could be genuine halo stars on extreme Galactic orbits, we considered the relevance of three ejection scenarios for our program stars, that is, the Hills scenario, the binary supernova scenario, and a potential extragalactic origin.
The Hills slingshot scenario may be valid only for two of our program stars because their last disk passages came close to the GC (J1632 and J1644).
The lifetime of EHB stars is about 100 Myr.
If the stars have been formed in a binary and then have been disrupted by the SMBH, the travel time from the GC to their current position must be consistent with this lifetime.
This is the case for J1632.
For J1644 this is only the case if we adopt the weighted mean of the catalogue proper motions, however.
When we adopt the proper motion we measured on our own, the travel
time is far too long, which means, if this scenario is true, that the star must have evolved to an sdB after the former binary was disrupted by the SMBH through one of the single evolution channels for hot subdwarfs.
Another option is the disruption of a hierarchical triple by the SMBH and the subsequent production of an sdB through the merger of two helium white dwarfs.
Alternatively, the star has evolved to an sdB with a low-mass companion, such as~a planet, that probably did not survive the common-envelope phase.
Accurate astrometry by Gaia will solve this uncertainty in the proper motion measurements. 
Figure \ref{fig_kine_J1644_Gaia} shows how the area of disk passages shrinks when the uncertainties in proper motion are reduced. An uncertainty of 0.1\,mas\,yr$^{-1}$ was also applied, which is a realistic uncertainty that the Gaia mission will provide \citepads{2012Ap&SS.341...31D}.

As the star with the highest velocity known (US 708, \citeads{2015Sci...347.1126G}) is a hot subdwarf that was not accelerated by the slingshot mechanism but rather a supernova explosion in a close binary, this is the scenario that should be considered next.
J2050 is a spectroscopic twin of US 708 and therefore a promising candidate of a surviving secondary of a supernova, as proposed for US 708.
It could be originating from a system similar to CD--30$^\circ$ 11223 \citepads{2013A&A...554A..54G}.
For J2050 the ejection velocity $\varv_\text{ej}=385\pm79\,\text{km\,s}^{-1}$ and the $\varv_\text{rot}\sin i<38\,\text{km\,s}^{-1}$  are both
moderate in comparison to the values of US 708: $\varv_\text{ej}=998\pm68\,\text{km\,s}^{-1}$, $\varv_\text{rot}\sin i=115\pm8\,\text{km\,s}^{-1}$.
As subdwarfs in compact binaries are assumed to have been spun up by the tidal influence, the progenitor system does not need to have been as tight as CD--30$^\circ$ 11223.
The progenitor system of J2050 could have had properties similar to that of the sdB + WD binary KPD 1930+2752 (\citeads{2000MNRAS.317L..41M}; \citeads{2007A&A...464..299G}).
In this system, the time in which the two objects will merge as a result of the radiation of gravitational waves is about twice as long as the lifetime of the sdB on the EHB.
Owing to the shrinkage of the orbit, Roche-lobe overflow might be possible before the sdB evolves into a white dwarf \citepads{2007A&A...464..299G}.
The travel time from the disk to the current position of J2050 is consistent with the lifetime.
The same is true for the potential disk runaway sdB star J1231.

The accretion scenario has been proposed by \citetads{2016ApJ...821L..13N} in order to explain the origin of the binary sdB J1211. 
According to this, J1211 was accreted from the debris of a destroyed satellite galaxy.
This scenario could also be valid for our program stars. 
If this is the case, the stars should belong to stellar streams in the halo that are yet to be discovered from Gaia astrometry.

\begin{acknowledgements}
E.Z. acknowledges funding by the German Science foundation (DFG) through grant HE1356/45-2. 
We thank John E.~Davis for the development of the {\sc slxfig} module that we used to prepare the figures in this paper. 
Some of the data presented in this paper were obtained from the Mikulski Archive for Space Telescopes (MAST). 
STScI is operated by the Association of Universities for Research in Astronomy, Inc., under NASA contract NAS5-26555. Support for MAST for non-HST data is provided by the NASA Office of Space Science via grant NNX09AF08G and by other grants and contracts. 
This work is based in part on data obtained as part of the UKIRT Infrared Deep Sky Survey. 
This publication makes use of data products from the Wide-field Infrared Survey Explorer, which is a joint project of the University of California, Los Angeles, and the Jet Propulsion Laboratory/California Institute of Technology, funded by the National Aeronautics and Space Administration. 
Based on observations collected at the European Organisation for Astronomical Research in the Southern Hemisphere under ESO program 093.D-0127(A).
This work is based on observations obtained at the W.M.~Keck Observatory, which is operated as a scientific partnership among the California Institute of Technology, the University of California, and the National Aeronautics and Space Administration.
The Observatory was made possible by the generous financial support of the W.M.~Keck Foundation. 
The authors wish to recognize and acknowledge the very significant cultural role and reverence that the summit of Maunakea has always had within the indigenous Hawaiian community.
We are most fortunate to have the opportunity to conduct observations from this mountain.
\end{acknowledgements}

\bibliography{sdb} 

\begin{thebibliography}{74}
\expandafter\ifx\csname natexlab\endcsname\relax\def\natexlab#1{#1}\fi

\bibitem[{{Abadi} {et~al.}(2009){Abadi}, {Navarro}, \&
  {Steinmetz}}]{2009ApJ...691L..63A}
{Abadi}, M.~G., {Navarro}, J.~F., \& {Steinmetz}, M. 2009, \apjl, 691, L63

\bibitem[{{Ahn} {et~al.}(2012){Ahn}, {Alexandroff}, {Allende Prieto},
  {Anderson}, {Anderton}, {Andrews}, {Aubourg}, {Bailey}, {Balbinot}, {Barnes},
  \& et~al.}]{2012ApJS..203...21A}
{Ahn}, C.~P., {Alexandroff}, R., {Allende Prieto}, C., {et~al.} 2012, \apjs,
  203, 21

\bibitem[{{Altmann} {et~al.}(2017){Altmann}, {Roeser}, {Demleitner}, {Bastian},
  \& {Schilbach}}]{2017arXiv170102629A}
{Altmann}, M., {Roeser}, S., {Demleitner}, M., {Bastian}, U., \& {Schilbach},
  E. 2017, ArXiv e-prints

\bibitem[{{Barlow} {et~al.}(2013){Barlow}, {Liss}, {Wade}, \&
  {Green}}]{2013ApJ...771...23B}
{Barlow}, B.~N., {Liss}, S.~E., {Wade}, R.~A., \& {Green}, E.~M. 2013, \apj,
  771, 23

\bibitem[{{Blaauw}(1961)}]{1961BAN....15..265B}
{Blaauw}, A. 1961, \bain, 15, 265

\bibitem[{{Brown}(2015)}]{2015ARA&A..53...15B}
{Brown}, W.~R. 2015, \araa, 53, 15

\bibitem[{{Brown} {et~al.}(2015){Brown}, {Anderson}, {Gnedin}, {Bond},
  {Geller}, \& {Kenyon}}]{2015ApJ...804...49B}
{Brown}, W.~R., {Anderson}, J., {Gnedin}, O.~Y., {et~al.} 2015, \apj, 804, 49

\bibitem[{{Brown} {et~al.}(2010){Brown}, {Anderson}, {Gnedin}, {Bond},
  {Geller}, {Kenyon}, \& {Livio}}]{2010ApJ...719L..23B}
{Brown}, W.~R., {Anderson}, J., {Gnedin}, O.~Y., {et~al.} 2010, \apjl, 719, L23

\bibitem[{{Brown} {et~al.}(2012){Brown}, {Cohen}, {Geller}, \&
  {Kenyon}}]{2012ApJ...754L...2B}
{Brown}, W.~R., {Cohen}, J.~G., {Geller}, M.~J., \& {Kenyon}, S.~J. 2012,
  \apjl, 754, L2

\bibitem[{{Brown} {et~al.}(2014){Brown}, {Geller}, \&
  {Kenyon}}]{2014ApJ...787...89B}
{Brown}, W.~R., {Geller}, M.~J., \& {Kenyon}, S.~J. 2014, \apj, 787, 89

\bibitem[{{Brown} {et~al.}(2005){Brown}, {Geller}, {Kenyon}, \&
  {Kurtz}}]{2005ApJ...622L..33B}
{Brown}, W.~R., {Geller}, M.~J., {Kenyon}, S.~J., \& {Kurtz}, M.~J. 2005,
  \apjl, 622, L33

\bibitem[{{Brown} {et~al.}(2007){Brown}, {Geller}, {Kenyon}, {Kurtz}, \&
  {Bromley}}]{2007ApJ...671.1708B}
{Brown}, W.~R., {Geller}, M.~J., {Kenyon}, S.~J., {Kurtz}, M.~J., \& {Bromley},
  B.~C. 2007, \apj, 671, 1708

\bibitem[{{Copperwheat} {et~al.}(2011){Copperwheat}, {Morales-Rueda}, {Marsh},
  {Maxted}, \& {Heber}}]{2011MNRAS.415.1381C}
{Copperwheat}, C.~M., {Morales-Rueda}, L., {Marsh}, T.~R., {Maxted}, P.~F.~L.,
  \& {Heber}, U. 2011, \mnras, 415, 1381

\bibitem[{{Cutri} \& {et al.}(2013)}]{2013yCat.2328....0C}
{Cutri}, R.~M. \& {et al.} 2013, VizieR Online Data Catalog, 2328

\bibitem[{{de Bruijne}(2012)}]{2012Ap&SS.341...31D}
{de Bruijne}, J.~H.~J. 2012, \apss, 341, 31

\bibitem[{{Edelmann} {et~al.}(2003){Edelmann}, {Heber}, {Hagen}, {Lemke},
  {Dreizler}, {Napiwotzki}, \& {Engels}}]{2003A&A...400..939E}
{Edelmann}, H., {Heber}, U., {Hagen}, H.-J., {et~al.} 2003, \aap, 400, 939

\bibitem[{{Edelmann} {et~al.}(2005){Edelmann}, {Napiwotzki}, {Heber},
  {Christlieb}, \& {Reimers}}]{2005ApJ...634L.181E}
{Edelmann}, H., {Napiwotzki}, R., {Heber}, U., {Christlieb}, N., \& {Reimers},
  D. 2005, \apjl, 634, L181

\bibitem[{{Fitzpatrick}(1999)}]{1999PASP..111...63F}
{Fitzpatrick}, E.~L. 1999, \pasp, 111, 63

\bibitem[{{Ganeshalingam} {et~al.}(2011){Ganeshalingam}, {Li}, \&
  {Filippenko}}]{2011MNRAS.416.2607G}
{Ganeshalingam}, M., {Li}, W., \& {Filippenko}, A.~V. 2011, \mnras, 416, 2607

\bibitem[{{Geier} {et~al.}(2015{\natexlab{a}}){Geier}, {F{\"u}rst}, {Ziegerer},
  {Kupfer}, {Heber}, {Irrgang}, {Wang}, {Liu}, {Han}, {Sesar}, {Levitan},
  {Kotak}, {Magnier}, {Smith}, {Burgett}, {Chambers}, {Flewelling}, {Kaiser},
  {Wainscoat}, \& {Waters}}]{2015Sci...347.1126G}
{Geier}, S., {F{\"u}rst}, F., {Ziegerer}, E., {et~al.} 2015{\natexlab{a}},
  Science, 347, 1126

\bibitem[{{Geier} \& {Heber}(2012)}]{2012A&A...543A.149G}
{Geier}, S. \& {Heber}, U. 2012, \aap, 543, A149

\bibitem[{{Geier} {et~al.}(2015{\natexlab{b}}){Geier}, {Kupfer}, {Heber},
  {Schaffenroth}, {Barlow}, {{\O}stensen}, {O'Toole}, {Ziegerer}, {Heuser},
  {Maxted}, {G{\"a}nsicke}, {Marsh}, {Napiwotzki}, {Br{\"u}nner},
  {Schindewolf}, \& {Niederhofer}}]{2015A&A...577A..26G}
{Geier}, S., {Kupfer}, T., {Heber}, U., {et~al.} 2015{\natexlab{b}}, \aap, 577,
  A26

\bibitem[{{Geier} {et~al.}(2013){Geier}, {Marsh}, {Wang}, {Dunlap}, {Barlow},
  {Schaffenroth}, {Chen}, {Irrgang}, {Maxted}, {Ziegerer}, {Kupfer},
  {Miszalski}, {Heber}, {Han}, {Shporer}, {Telting}, {G{\"a}nsicke},
  {{\O}stensen}, {O'Toole}, \& {Napiwotzki}}]{2013A&A...554A..54G}
{Geier}, S., {Marsh}, T.~R., {Wang}, B., {et~al.} 2013, \aap, 554, A54

\bibitem[{{Geier} {et~al.}(2007){Geier}, {Nesslinger}, {Heber}, {Przybilla},
  {Napiwotzki}, \& {Kudritzki}}]{2007A&A...464..299G}
{Geier}, S., {Nesslinger}, S., {Heber}, U., {et~al.} 2007, \aap, 464, 299

\bibitem[{{Ghez} {et~al.}(2005){Ghez}, {Salim}, {Hornstein}, {Tanner}, {Lu},
  {Morris}, {Becklin}, \& {Duch{\^e}ne}}]{2005ApJ...620..744G}
{Ghez}, A.~M., {Salim}, S., {Hornstein}, S.~D., {et~al.} 2005, \apj, 620, 744

\bibitem[{{Gillessen} {et~al.}(2009){Gillessen}, {Eisenhauer}, {Trippe},
  {Alexander}, {Genzel}, {Martins}, \& {Ott}}]{2009ApJ...692.1075G}
{Gillessen}, S., {Eisenhauer}, F., {Trippe}, S., {et~al.} 2009, \apj, 692, 1075

\bibitem[{{Han}(2008)}]{2008A&A...484L..31H}
{Han}, Z. 2008, \aap, 484, L31

\bibitem[{{Heber}(2009)}]{2009ARA&A..47..211H}
{Heber}, U. 2009, \araa, 47, 211

\bibitem[{{Heber}(2016)}]{2016PASP..128h2001H}
{Heber}, U. 2016, \pasp, 128, 082001

\bibitem[{{Heber} {et~al.}(2008){Heber}, {Edelmann}, {Napiwotzki}, {Altmann},
  \& {Scholz}}]{2008A&A...483L..21H}
{Heber}, U., {Edelmann}, H., {Napiwotzki}, R., {Altmann}, M., \& {Scholz},
  R.-D. 2008, \aap, 483, L21

\bibitem[{{Heber} {et~al.}(2000){Heber}, {Reid}, \&
  {Werner}}]{2000A&A...363..198H}
{Heber}, U., {Reid}, I.~N., \& {Werner}, K. 2000, \aap, 363, 198

\bibitem[{{Hernquist}(1990)}]{1990ApJ...356..359H}
{Hernquist}, L. 1990, \apj, 356, 359

\bibitem[{{Hills}(1988)}]{1988Natur.331..687H}
{Hills}, J.~G. 1988, \nat, 331, 687

\bibitem[{{Hirsch} \& {Heber}(2009)}]{2009JPhCS.172a2015H}
{Hirsch}, H. \& {Heber}, U. 2009, Journal of Physics Conference Series, 172,
  012015

\bibitem[{{Hirsch}(2009)}]{2009PhDT.......273H}
{Hirsch}, H.~A. 2009, PhD thesis, Friedrich-Alexander University
  Erlangen-N{\"u}rnberg

\bibitem[{{Hirsch} {et~al.}(2005){Hirsch}, {Heber}, {O'Toole}, \&
  {Bresolin}}]{2005A&A...444L..61H}
{Hirsch}, H.~A., {Heber}, U., {O'Toole}, S.~J., \& {Bresolin}, F. 2005, \aap,
  444, L61

\bibitem[{{Irrgang} {et~al.}(2010){Irrgang}, {Przybilla}, {Heber}, {Nieva}, \&
  {Schuh}}]{2010ApJ...711..138I}
{Irrgang}, A., {Przybilla}, N., {Heber}, U., {Nieva}, M.~F., \& {Schuh}, S.
  2010, \apj, 711, 138

\bibitem[{{Irrgang} {et~al.}(2013){Irrgang}, {Wilcox}, {Tucker}, \&
  {Schiefelbein}}]{2013A&A...549A.137I}
{Irrgang}, A., {Wilcox}, B., {Tucker}, E., \& {Schiefelbein}, L. 2013, \aap,
  549, A137

\bibitem[{{Kurucz}(1996)}]{1996ASPC..108..160K}
{Kurucz}, R.~L. 1996, in Model Atmospheres and Spectrum Synthesis, ed. S.~J.\
  {Adelman}, F.\ {Kupka}, \& W.~W.\ {Weiss} (San Francisco: ASP), 160

\bibitem[{{Lawrence} {et~al.}(2007){Lawrence}, {Warren}, {Almaini}, {Edge},
  {Hambly}, {Jameson}, {Lucas}, {Casali}, {Adamson}, {Dye}, {Emerson},
  {Foucaud}, {Hewett}, {Hirst}, {Hodgkin}, {Irwin}, {Lodieu}, {McMahon},
  {Simpson}, {Smail}, {Mortlock}, \& {Folger}}]{2007MNRAS.379.1599L}
{Lawrence}, A., {Warren}, S.~J., {Almaini}, O., {et~al.} 2007, \mnras, 379,
  1599

\bibitem[{{Leonard}(1991)}]{1991AJ....101..562L}
{Leonard}, P.~J.~T. 1991, \aj, 101, 562

\bibitem[{{Lisker} {et~al.}(2005){Lisker}, {Heber}, {Napiwotzki}, {Christlieb},
  {Han}, {Homeier}, \& {Reimers}}]{2005A&A...430..223L}
{Lisker}, T., {Heber}, U., {Napiwotzki}, R., {et~al.} 2005, \aap, 430, 223

\bibitem[{{Madigan} {et~al.}(2014){Madigan}, {Pfuhl}, {Levin}, {Gillessen},
  {Genzel}, \& {Perets}}]{2014ApJ...784...23M}
{Madigan}, A.-M., {Pfuhl}, O., {Levin}, Y., {et~al.} 2014, \apj, 784, 23

\bibitem[{{Maxted} {et~al.}(2001){Maxted}, {Heber}, {Marsh}, \&
  {North}}]{2001MNRAS.326.1391M}
{Maxted}, P.~F.~L., {Heber}, U., {Marsh}, T.~R., \& {North}, R.~C. 2001,
  \mnras, 326, 1391

\bibitem[{{Maxted} {et~al.}(2000){Maxted}, {Marsh}, \&
  {North}}]{2000MNRAS.317L..41M}
{Maxted}, P.~F.~L., {Marsh}, T.~R., \& {North}, R.~C. 2000, \mnras, 317, L41

\bibitem[{{Miyamoto} \& {Nagai}(1975)}]{1975PASJ...27..533M}
{Miyamoto}, M. \& {Nagai}, R. 1975, \pasj, 27, 533

\bibitem[{{Monet} {et~al.}(2003){Monet}, {Levine}, {Canzian}, {Ables}, {Bird},
  {Dahn}, {Guetter}, {Harris}, {Henden}, {Leggett}, {Levison}, {Luginbuhl},
  {Martini}, {Monet}, {Munn}, {Pier}, {Rhodes}, {Riepe}, {Sell}, {Stone},
  {Vrba}, {Walker}, {Westerhout}, {Brucato}, {Reid}, {Schoening}, {Hartley},
  {Read}, \& {Tritton}}]{2003AJ....125..984M}
{Monet}, D.~G., {Levine}, S.~E., {Canzian}, B., {et~al.} 2003, \aj, 125, 984

\bibitem[{{Morales-Rueda} {et~al.}(2003){Morales-Rueda}, {Maxted}, {Marsh},
  {North}, \& {Heber}}]{2003MNRAS.338..752M}
{Morales-Rueda}, L., {Maxted}, P.~F.~L., {Marsh}, T.~R., {North}, R.~C., \&
  {Heber}, U. 2003, \mnras, 338, 752

\bibitem[{{Napiwotzki} {et~al.}(2004){Napiwotzki}, {Karl}, {Lisker}, {Heber},
  {Christlieb}, {Reimers}, {Nelemans}, \& {Homeier}}]{2004Ap&SS.291..321N}
{Napiwotzki}, R., {Karl}, C.~A., {Lisker}, T., {et~al.} 2004, \apss, 291, 321

\bibitem[{{Naslim} {et~al.}(2013){Naslim}, {Jeffery}, {Hibbert}, \&
  {Behara}}]{2013MNRAS.434.1920N}
{Naslim}, N., {Jeffery}, C.~S., {Hibbert}, A., \& {Behara}, N.~T. 2013, \mnras,
  434, 1920

\bibitem[{{Navarro} {et~al.}(1997){Navarro}, {Frenk}, \&
  {White}}]{1997ApJ...490..493N}
{Navarro}, J.~F., {Frenk}, C.~S., \& {White}, S.~D.~M. 1997, \apj, 490, 493

\bibitem[{{N{\'e}meth} {et~al.}(2016){N{\'e}meth}, {Ziegerer}, {Irrgang},
  {Geier}, {F{\"u}rst}, {Kupfer}, \& {Heber}}]{2016ApJ...821L..13N}
{N{\'e}meth}, P., {Ziegerer}, E., {Irrgang}, A., {et~al.} 2016, \apjl, 821, L13

\bibitem[{{O'Toole} \& {Heber}(2006)}]{2006A&A...452..579O}
{O'Toole}, S.~J. \& {Heber}, U. 2006, \aap, 452, 579

\bibitem[{{Pan} {et~al.}(2013){Pan}, {Ricker}, \& {Taam}}]{2013ApJ...773...49P}
{Pan}, K.-C., {Ricker}, P.~M., \& {Taam}, R.~E. 2013, \apj, 773, 49

\bibitem[{{Pauli} {et~al.}(2006){Pauli}, {Napiwotzki}, {Heber}, {Altmann}, \&
  {Odenkirchen}}]{2006A&A...447..173P}
{Pauli}, E.-M., {Napiwotzki}, R., {Heber}, U., {Altmann}, M., \& {Odenkirchen},
  M. 2006, \aap, 447, 173

\bibitem[{{Perets} \& {{\v S}ubr}(2012)}]{2012ApJ...751..133P}
{Perets}, H.~B. \& {{\v S}ubr}, L. 2012, \apj, 751, 133

\bibitem[{{Perets} {et~al.}(2009){Perets}, {Wu}, {Zhao}, {Famaey}, {Gentile},
  \& {Alexander}}]{2009ApJ...697.2096P}
{Perets}, H.~B., {Wu}, X., {Zhao}, H.~S., {et~al.} 2009, \apj, 697, 2096

\bibitem[{{Przybilla} {et~al.}(2008{\natexlab{a}}){Przybilla}, {Fernanda
  Nieva}, {Heber}, \& {Butler}}]{2008ApJ...684L.103P}
{Przybilla}, N., {Fernanda Nieva}, M., {Heber}, U., \& {Butler}, K.
  2008{\natexlab{a}}, \apjl, 684, L103

\bibitem[{{Przybilla} {et~al.}(2008{\natexlab{b}}){Przybilla}, {Nieva},
  {Heber}, {Firnstein}, {Butler}, {Napiwotzki}, \&
  {Edelmann}}]{2008A&A...480L..37P}
{Przybilla}, N., {Nieva}, M.~F., {Heber}, U., {et~al.} 2008{\natexlab{b}},
  \aap, 480, L37

\bibitem[{{Qi} {et~al.}(2015){Qi}, {Yu}, {Bucciarelli}, {Lattanzi}, {Smart},
  {Spagna}, {McLean}, {Tang}, {Jones}, {Morbidelli}, {Nicastro}, \&
  {Vecchiato}}]{2015AJ....150..137Q}
{Qi}, Z., {Yu}, Y., {Bucciarelli}, B., {et~al.} 2015, \aj, 150, 137

\bibitem[{{Ramspeck} {et~al.}(2001){Ramspeck}, {Heber}, \&
  {Moehler}}]{2001A&A...378..907R}
{Ramspeck}, M., {Heber}, U., \& {Moehler}, S. 2001, \aap, 378, 907

\bibitem[{{Randall} {et~al.}(2015){Randall}, {Bagnulo}, {Ziegerer}, {Geier}, \&
  {Fontaine}}]{2015A&A...576A..65R}
{Randall}, S.~K., {Bagnulo}, S., {Ziegerer}, E., {Geier}, S., \& {Fontaine}, G.
  2015, \aap, 576, A65

\bibitem[{{Roeser} {et~al.}(2010){Roeser}, {Demleitner}, \&
  {Schilbach}}]{2010AJ....139.2440R}
{Roeser}, S., {Demleitner}, M., \& {Schilbach}, E. 2010, \aj, 139, 2440

\bibitem[{{Rossi} {et~al.}(2017){Rossi}, {Marchetti}, {Cacciato}, {Kuiack}, \&
  {Sari}}]{2017MNRAS.tmp..102R}
{Rossi}, E.~M., {Marchetti}, T., {Cacciato}, M., {Kuiack}, M., \& {Sari}, R.
  2017, \mnras

\bibitem[{{Schaffenroth} {et~al.}(2015){Schaffenroth}, {Barlow}, {Drechsel}, \&
  {Dunlap}}]{2015A&A...576A.123S}
{Schaffenroth}, V., {Barlow}, B.~N., {Drechsel}, H., \& {Dunlap}, B.~H. 2015,
  \aap, 576, A123

\bibitem[{{Sch{\"o}del} {et~al.}(2003){Sch{\"o}del}, {Ott}, {Genzel}, {Eckart},
  {Mouawad}, \& {Alexander}}]{2003ApJ...596.1015S}
{Sch{\"o}del}, R., {Ott}, T., {Genzel}, R., {et~al.} 2003, \apj, 596, 1015

\bibitem[{{Smith} {et~al.}(2009){Smith}, {Ebeling}, {Limousin}, {Kneib},
  {Swinbank}, {Ma}, {Jauzac}, {Richard}, {Jullo}, {Sand}, {Edge}, \&
  {Smail}}]{2009ApJ...707L.163S}
{Smith}, G.~P., {Ebeling}, H., {Limousin}, M., {et~al.} 2009, \apjl, 707, L163

\bibitem[{{Stroeer} {et~al.}(2007){Stroeer}, {Heber}, {Lisker}, {Napiwotzki},
  {Dreizler}, {Christlieb}, \& {Reimers}}]{2007A&A...462..269S}
{Stroeer}, A., {Heber}, U., {Lisker}, T., {et~al.} 2007, \aap, 462, 269

\bibitem[{{Svensson} {et~al.}(2008){Svensson}, {Church}, \&
  {Davies}}]{2008MNRAS.383L..15S}
{Svensson}, K.~M., {Church}, R.~P., \& {Davies}, M.~B. 2008, \mnras, 383, L15

\bibitem[{{Tauris} \& {Takens}(1998)}]{1998A&A...330.1047T}
{Tauris}, T.~M. \& {Takens}, R.~J. 1998, \aap, 330, 1047

\bibitem[{{Tillich} {et~al.}(2011){Tillich}, {Heber}, {Geier}, {Hirsch},
  {Maxted}, {G{\"a}nsicke}, {Marsh}, {Napiwotzki}, {{\O}stensen}, \&
  {Scholz}}]{2011A&A...527A.137T}
{Tillich}, A., {Heber}, U., {Geier}, S., {et~al.} 2011, \aap, 527, A137

\bibitem[{{Vos} {et~al.}(2012){Vos}, {{\O}stensen}, {Degroote}, {De Smedt},
  {Green}, {Heber}, {Van Winckel}, {Acke}, {Bloemen}, {De Cat}, {Exter},
  {Lampens}, {Lombaert}, {Masseron}, {Menu}, {Neyskens}, {Raskin}, {Ringat},
  {Rauch}, {Smolders}, \& {Tkachenko}}]{2012A&A...548A...6V}
{Vos}, J., {{\O}stensen}, R.~H., {Degroote}, P., {et~al.} 2012, \aap, 548, A6

\bibitem[{{Vos} {et~al.}(2013){Vos}, {{\O}stensen}, {N{\'e}meth}, {Green},
  {Heber}, \& {Van Winckel}}]{2013A&A...559A..54V}
{Vos}, J., {{\O}stensen}, R.~H., {N{\'e}meth}, P., {et~al.} 2013, \aap, 559,
  A54

\bibitem[{{Zacharias} {et~al.}(2012){Zacharias}, {Finch}, {Girard}, {Henden},
  {Bartlett}, {Monet}, \& {Zacharias}}]{2012yCat.1322....0Z}
{Zacharias}, N., {Finch}, C.~T., {Girard}, T.~M., {et~al.} 2012, VizieR Online
  Data Catalog, 1322

\end{thebibliography}
\bibliographystyle{aa}

\begin{appendix}
\section{}

\begin{table*}
\caption{\label{tab_spectra} Atmospheric parameters, radial velocities, and projected rotational velocities}
\renewcommand{\arraystretch}{1.4}
\begin{tabular}{llcccccc}
\hline\hline
Name & OBS & $T_\text{eff}$ & $\log g$ & $\log\frac{n(\text{He})}{n(\text{H})}$ & $\varv_\text{rot}\sin i$ & $\varv_\text{rad}$ \\
 & & (K) & (cgs) & & (km/s) & (km/s) \\
\hline
J1231 & SDSS-BOSS & $25200\pm400$ & $5.11\pm0.04$ & $-2.29\pm0.16$ & $$ & $460\pm8$ \\
 & XSHOOTER/VLT\tablefootmark{b} & $25200\pm500$ & $5.13\pm0.02$ & $-2.23\pm0.05$ & $<45$ & $467\pm2$ \\
J1632 & SDSS\tablefootmark{a} & $26870\pm610$ & $5.31\pm0.09$ & $-2.1\pm0.2$ &  & $-239\pm10$ \\
 & SDSS-BOSS & $29000\pm500$ & $5.46\pm0.06$ & $-1.59\pm0.09$ & $$ & $-261\pm20$ \\
 & ESI/Keck & $29500\pm400$ & $5.61\pm0.07$ & $-1.78\pm0.04$ & $<35$ & $-253\pm10$\\
 & XSHOOTER/VLT\tablefootmark{b} & $28900\pm500$ & $5.61\pm0.02$ & $-1.83\pm0.03$ & $<33$ & $-239\pm2$ \\
J1644 & SDSS\tablefootmark{a} & $31680\pm410$ & $5.78\pm0.11$ & $-2.9\pm0.3$ & $$ & $-314\pm5$ \\
 & SDSS-BOSS\tablefootmark{b} & $33400\pm200$ & $5.69\pm0.04$ & $<3.0$ & $$ & $-309\pm9$ \\
 & ESI/Keck\tablefootmark{b} & $33800\pm200$ & $5.76\pm0.04$ & $<-3.0$ & $<38$ & $-299\pm10$ \\
J2050 & SDSS\tablefootmark{b} & $48000\pm500$ & $5.68\pm0.05$ & $>+1.3$ &$$ & $-509\pm19$ \\
 & FORS1/VLT & $48600\pm700$ & $5.84\pm0.12$ & $>+2.0$ & $$ & $-485\pm44$ \\
 & ESI/Keck\tablefootmark{b} & $47000\pm200$ & $5.71\pm0.06$ & $>+2.0$ & $<38$ & $-473\pm10$ \\
\hline
\end{tabular}                                                  
\centering
\tablefoot{SDSS-BOSS: $R=2200$, 3600-10000\AA, ESI: echellette mode with 0.5 arcsec-slit, $R=8000$, 4000-6000\AA, XSHOOTER: $R=10000$, 3000-6800\AA, FORS1: $R=1800$, 3730-5200\AA. \tablefoottext{a}{Values are taken from \citetads{2011A&A...527A.137T}.}\tablefoottext{b}{Values adopted for the kinematic calculations in this paper.}}
\end{table*}

\end{appendix}
  
\end{document}